\documentclass{elsart}
\usepackage{lineno}
\usepackage{epsfig}
\journal{Nuclear Instruments and Methods A}
\begin{document}
\begin{frontmatter}
\title{Radiation Hardness of Thin Low Gain Avalanche Detectors\thanksref{RD50}}
\thanks[RD50]{Work performed in the framework of the CERN-RD50 collaboration.}
\author{G.\,Kramberger$^{a}$, M. Carulla$^{b}$,  E. Cavallaro$^{c}$, V.\,Cindro$^{a}$, D. Flores$^{b}$,}
\author{Z. Galloway$^{e}$, S. Grinstein$^{c,d}$, S.\,Hidalgo$^{b}$, V.\, Fadeyev$^{e}$,J. Lange$^{c}$,}
\author{I.\,Mandi\' c$^{a}$, G. Medin$^{f}$, A. Merlos$^{b}$, F. McKinney-Martinez$^{e}$,  }
\author{M.\,Miku\v z$^{a,g}$, D.\,Quirion$^{b}$, G. Pellegrini$^{b}$, M. Petek$^{a}$, }
\author{H. F-W. Sadrozinski$^{e}$, A. Seiden$^{e}$, M. Zavrtanik$^{a}$}
\address {
$^a$ Jo\v zef Stefan Institute, Jamova 39, SI-1000 Ljubljana, Slovenia\\
$^b$ Centro Nacional de Microelectr\'{o}nica (IMB-CNM-CSIC), Barcelona 08193, Spain \\
$^c$ Institut de F\'{i}sica d'Altes Energies (IFAE), The Barcelona Institute of Science and Technology (BIST), 08193 Bellaterra (Barcelona), Spain \\
$^d$ Instituci\'{o} Catalana de Recerca i Estudis Avan\c{c}ats (ICREA), Pg. Llu\'{i}s Companys 23, 08010 Barcelona, Spain \\
$^e$ UCSC, Santa Cruz Institute for Particle Physics, Santa Cruz, CA 95064, USA.  \\
$^f$ University of Montenegro, Faculty of Natural Sciences and Mathematics, Podgorica, Montenegro \\
$^g$ University of Ljubljana, Faculty of Mathematics and Physics, Jadranska 19, SI-1000 Ljubljana, Slovenia \\}
\thanks[hvala]{Corresponding author; Address:
Jo\v zef Stefan Institute, Jamova 39, SI-1000 Ljubljana,
Slovenia. Tel: (+386) 1 477 3512, fax: (+386) 1 477 3166,
e-mail: Gregor.Kramberger@ijs.si}     
\begin{abstract}
Low Gain Avalanche Detectors (LGAD) are based on a n$^{++}$-p$^{+}$-p-p$^{++}$ structure where an appropriate doping of the multiplication 
layer (p$^+$) leads to high enough electric fields for impact ionization. Gain factors of few tens in charge 
 significantly improve the resolution of timing measurements, particularly for thin detectors,
where the timing performance was shown to be limited by Landau fluctuations. The main obstacle 
for their operation is the decrease of gain with irradiation, attributed to effective acceptor removal in the gain layer. Sets of thin 
sensors were produced by two different producers on different substrates, with different gain layer doping profiles and thicknesses (45, 50 and 80 $\mu$m). 
Their performance in terms of gain/collected charge and leakage current was compared before and after irradiation with neutrons and 
pions up to the equivalent fluences of $5\cdot10^{15}$ cm$^{-2}$. Transient Current Technique and charge collection measurements with LHC 
speed electronics were employed to characterize the detectors. The thin LGAD sensors were shown to perform much better than sensors of standard 
thickness ($\sim$300 $\mu$m) and offer larger charge collection with respect to detectors without gain layer for fluences $<2\cdot10^{15}$ cm$^{-2}$. Larger initial gain prolongs the beneficial performance of LGADs. 
Pions were found to be more damaging than neutrons at the same equivalent fluence, while no significant difference was found between different producers. At very high fluences and bias voltages
the gain appears due to deep acceptors in the bulk, hence also in thin standard detectors.
\end{abstract}
\end{frontmatter}
\section{Introduction}

\begin{figure}[!hbt]
\begin{center}
\epsfig{file=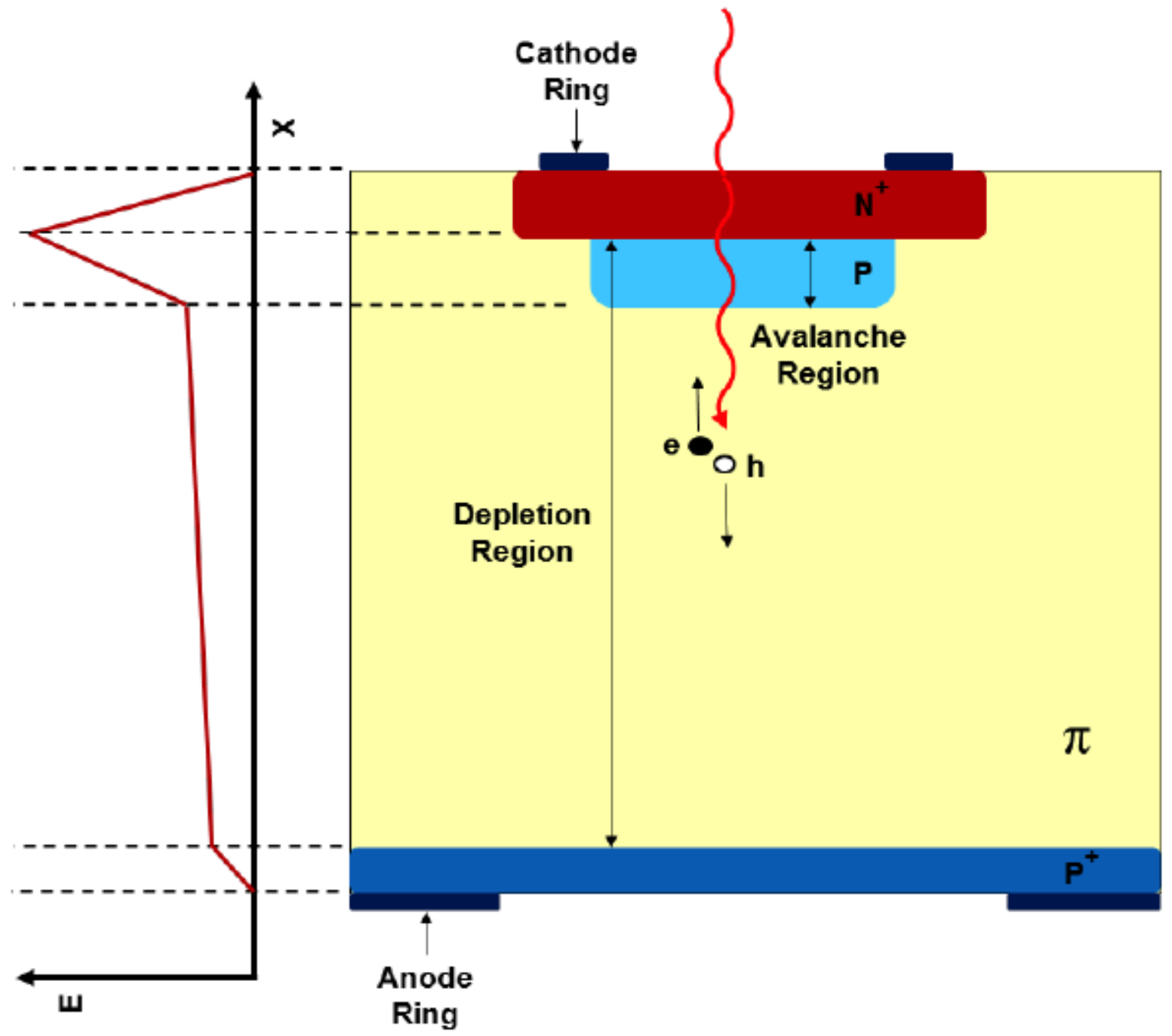,width=.4\linewidth,clip=}
\end{center}
\caption{Schematic view of the Low Gain Avalanche Detector. P-type substrate is denoted by $\pi$.}
\label{fi:LGAD}
\end{figure}

Low gain avalanche detectors (LGAD) exploit a n$^{++}$-p$^{+}$-p-p$^{++}$ structure to achieve high enough electric fields near the junction contact 
for impact ionization \cite{LGAD-Main} (see Fig. \ref{fi:LGAD}). The gain depends on the p$^+$ layer's doping level and profile shape. 
Usually gain factors of several tens were achieved in most LGADs produced so far. 
That assures efficient operation of thin sensors required for precise 
timing applications in particle physics \cite{LGAD-Thin}. A superb timing resolution of 26 ps per single LGAD layer was achieved recently \cite{LGAD-Timing},
which made thin LGADs ($\sim 50$ $\mu$m) a baseline option for timing detectors of both CMS and ATLAS at the high luminosity upgrade of the LHC (HL-LHC) around 2026 \cite{HGTD,CMS-TD}. 
The main obstacle for their successful use at future experiments in high energy physics is the degradation of gain with fluence. LGADs will be exposed 
at the HL-LHC to equivalent fluences of up to $\Phi_{eq}=6\cdot 10^{15}$ cm$^{-2}$. At these fluences the gain due 
to the p$^+$ layer completely disappears \cite{LGAD-radhard}. 

Excellent timing resolution can only be achieved if the induced current variations due to 
Landau fluctuations are minimized, therefore the use of thin devices is required \cite{NC2017}.

\begin{figure}[!hbt]
\begin{center}
\epsfig{file=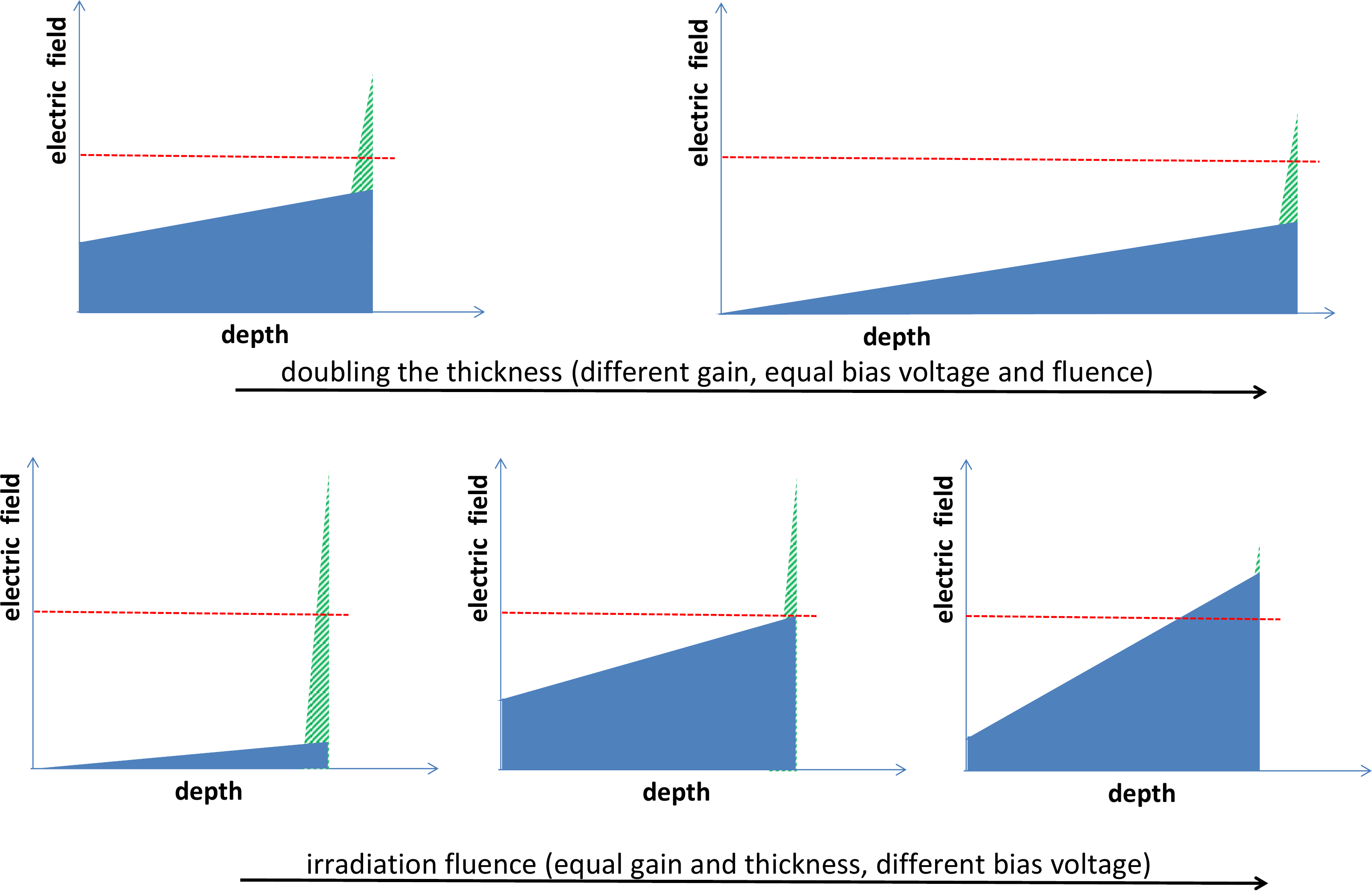,width=1.0\linewidth,clip=}
\end{center}
\caption{Schematic view of the electric field in irradiated LGAD detectors. The sharpening of the field due to shallow dopants is marked with pattern. The dashed red line denotes required electric field for sizable multiplication. }
\label{fi:EFieldSchematic}
\end{figure}

The use of thin devices also improves the radiation hardness as explained in the following.
After the device depletes the further increase of bias voltage roughly increases the electric
field by the same amount over the entire thickness of the device. This requires a much higher bias voltage 
in 300 $\mu$m thick LGAD detectors than in thin ones to retain high electric fields in the gain region. 
The device usually breaks down before the gain can be recovered. With an appropriate design a thin sensor tolerating a 
high bias voltage would therefore be more radiation hard, although only because of geometrical effect. This is illustrated in Fig. \ref{fi:EFieldSchematic}. 
At very high fluences the concentration of initial acceptors in the gain layer becomes negligible, but the deep 
acceptors created in the part of the bulk by irradiation lead to 
fields high enough for sizable multiplication as observed before \cite{JL2010,IM2009,CMGian}. 

In order to test this hypothesis and to establish the requirements needed for successful operation of thin LGAD sensors, sets of different 
thin LGADs produced by CNM\footnote{Centro Nacional de Microelectr\'{o}nica, Barcelona, Spain} and HPK \footnote{Hamamatsu Photonics,  Hamamatsu, Japan} 
were characterized before and after irradiations with different particles. The timing measurements with these LGADs are presented in other papers \cite{JL,HS}.
It was found that after low and moderate fluences ($<1-2\cdot 10^{15}$ cm$^{-2}$) the gain solely determines the timing resolution once the 
velocities close to saturation are achieved in the detector bulk (provided the noise is constant). Moreover at very high fluences the multiplication 
taking place in the bulk instead of in a narrow gain layer results in faster induced current rise, therefore improved timing resolution 
even at somewhat smaller gain. Establishing the relation between the collected charge and the required operation voltage in the complete fluence range of HL-LHC is 
therefore of utmost importance. While most of the recent irradiation studies concentrated on reactor neutrons, LGADs at HL-LHC will be exposed also 
to charged hadrons which were shown before \cite{LGAD-radhard} to cause faster effective acceptor removal at the same equivalent fluence. Therefore LGADs studied in the 
present work include also those irradiated by 200 MeV pions.

Comparison of devices of different thicknesses will also add to the understanding 
of the underlying mechanism of effective acceptor removal. The dependence of the effective acceptor removal rate on different producers will 
show its universality and possibly lead to improved doping profiles. Thin LGADs made on different substrates will also reveal their impact on 
charge collection.

\section{Samples and experimental technique}

\begin{figure}[!hbt]
\begin{center}
\epsfig{file=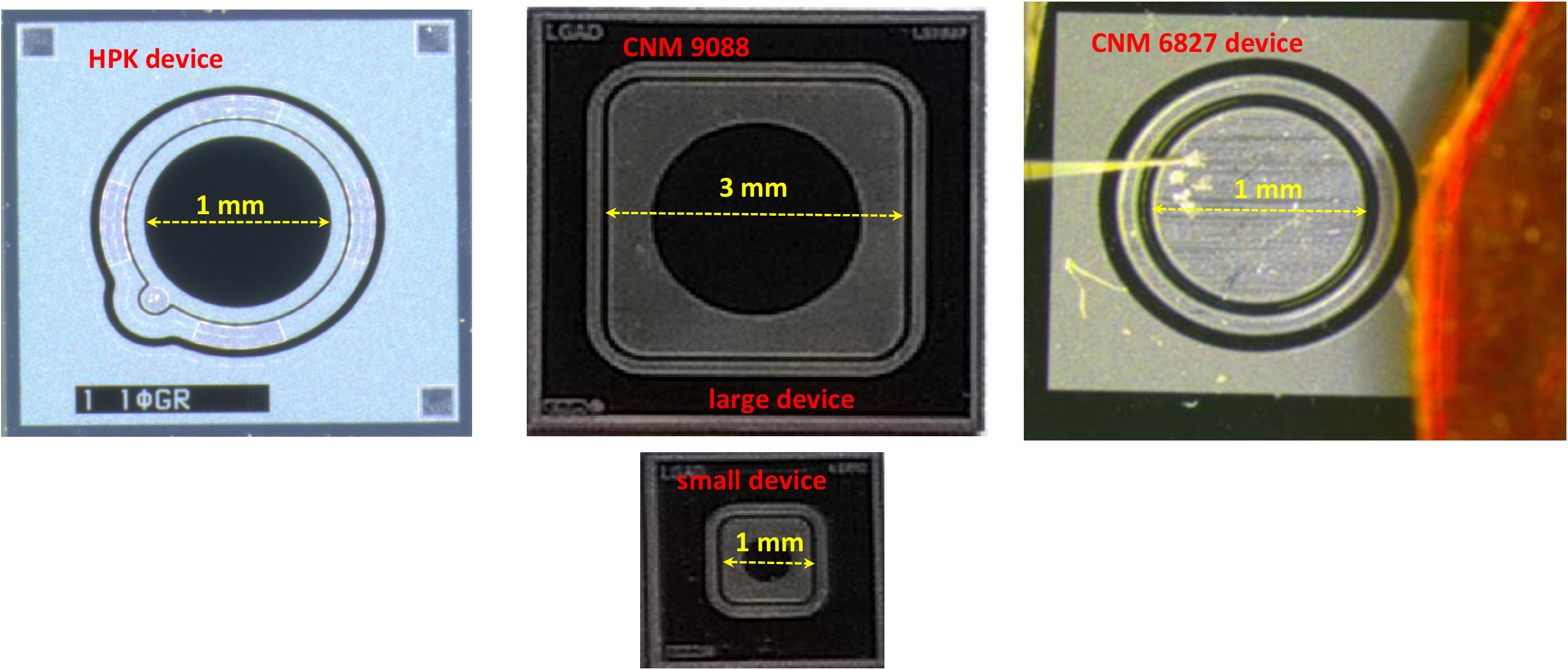,width=1.0\linewidth,clip=}
\end{center}
\caption{Devices used in the study.}
\label{fi:Devices}
\end{figure}

Three different sets of LGADs produced by CNM and HPK  were studied with their properties listed in Table \ref{ta:samples} and shown in Fig. \ref{fi:Devices}.
\begin{table}
\caption{
The sample names used in the text will be composed from Sample Set name (or producer) followed by size/thickness and dose indication. 
Samples from sets R9088 and ECX20840 had an opening in the metallization in the front contact allowing for light injection. Back 
contacts were fully metallized. The samples from CNM sets included also control samples (PIN), which were identical to corresponding LGADs without multiplication layer. 
}
\label{ta:samples}
\begin{center}
\begin{tabular}{|c|c|c|c|c|}
\hline
Sample set/Run & Producer & size-thickness & Implantation Dose & substrate  \\
\hline
R9088	& CNM & rectangular &  L = $1.8\cdot10^{13}$ cm$^{-2}$ & 300 $\mu$m SOI \\
	&     & L = $3\times3$ mm - 45 $\mu$m     & M = $1.9\cdot10^{13}$ cm$^{-2}$ & \\
	&     & S = $1.3\times1.3$ mm - 45 $\mu$m & H = $2.0\cdot10^{13}$  cm$^{-2}$ & \\
\hline
R6827	& CNM & circular $2R=1$ mm - 50 $\mu$m     &   $1.5\cdot10^{13}$ cm$^{-2}$ & 300 $\mu$m EPI \\
\hline
ECX20840& HPK &  circular $2R=1$ mm - 50 $\mu$m   & A,B,C,D  & 300 $\mu$m \\
or short&     &  circular $2R=1$ mm - 80 $\mu$m   & 4\% between splits &  low resistivity \\
HPK 	&     &                                   & A-lowest, D-highest & Si wafer\\
\hline
\end{tabular}
\end{center}
\end{table}
All three sets  were irradiated with neutrons at Jo\v zef 
Stefan Institute research reactor \cite{Reaktor1}. Set R9088 was irradiated also 
with 200 MeV pions at Paul Scherrer Institute in Villigen, Switzerland \cite{PSI}.
They were characterized by Transient Current Technique (TCT) \cite{VE1996} and charge collection efficiency  
measurement with electrons from a $^{90}$Sr source  with LHC speed electronics (CCM). The detailed description of the setups can be 
found in \cite{GK2002} for TCT and \cite{GK2005} for CCM. The very small size of the active pads required a small collimator (1 mm$^2$) 
which together with careful alignment assured that almost all the electrons that reach the scintillator and 
trigger the readout have crossed the detector. This allowed the measurements of the signal even 
at very low signal-to-noise (S/N) ratios. In some case for the smallest samples a non-perfect alignment was tolerated as the 
relatively large S/N for these devices allowed the separation of noise hits (i.e. electrons missing the active area and triggering the readout).

The measurements were performed after annealing for 80 min at 60$^{\circ}$C. Samples from sets HPK and R6827 
were irradiated in several steps  with the above mentioned annealing procedure
done after each step (CERN scenario \cite{RD48}). In this way it was possible to cover the fluence range with a limited number of samples. 
The fluences of particles were scaled to 1 MeV neutron equivalent
fluences by using hardness factors: 0.92 for reactor neutrons ($>100$ keV) \cite{Reaktor} and 1.14 for 200 MeV pions \cite{RD48}.

\section{Charge collection/gain}

Charge collection measurements were performed mostly at $T=-10^\circ$C. At each voltage point 2500 events ($^{90}$Sr electron triggers)
were recorded to disk and analyzed offline. The response of the preamplifier-shaping amplifier (25 ns peaking time) was calibrated with a  
non-irradiated standard n$^+$-p pad detector and also with 59.5 keV photons from a $^{241}$Am source, which allowed absolute charge measurements. 

\begin{figure}[!hbt]
\begin{center}
\epsfig{file=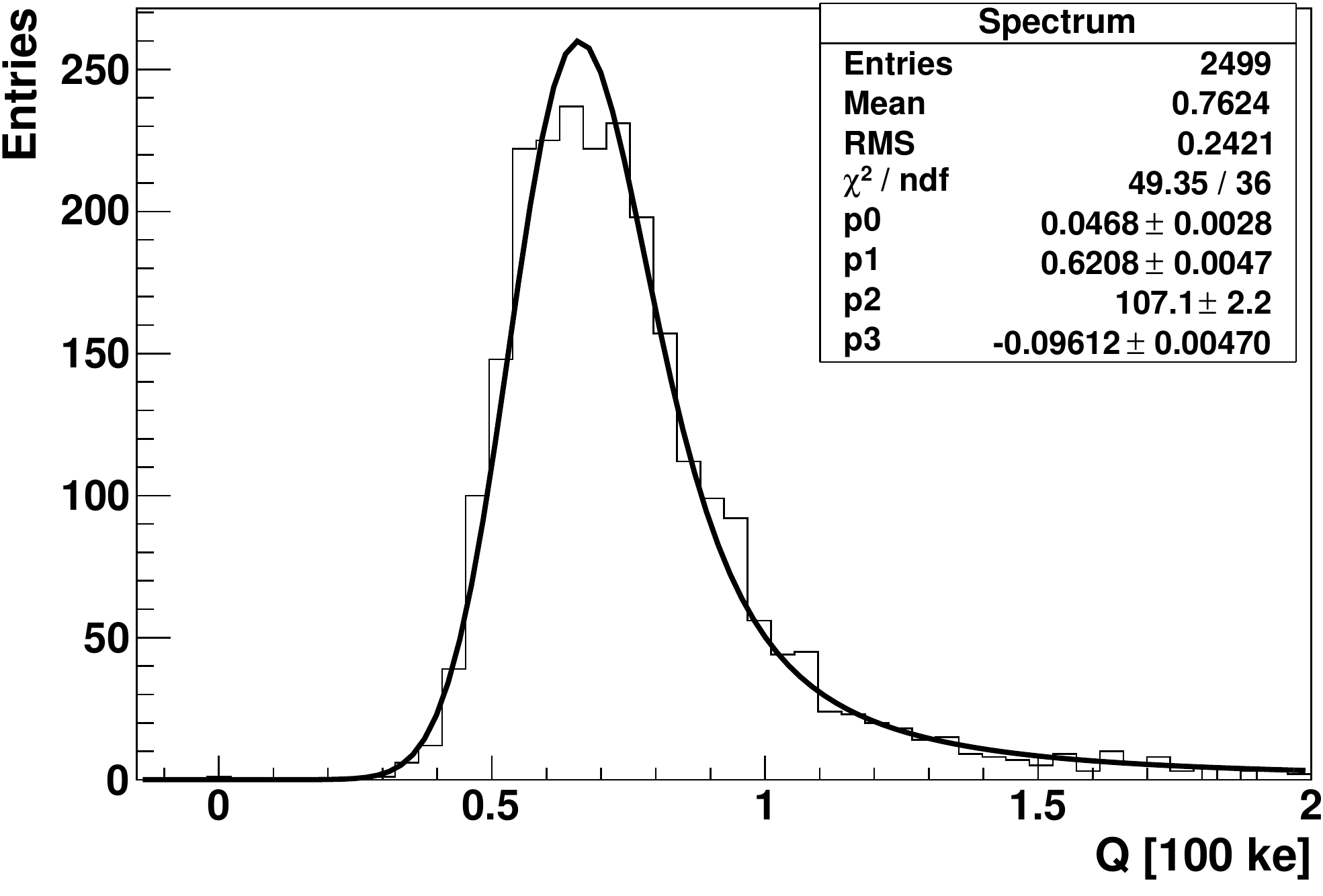,width=.5\linewidth,clip=}
\end{center}
\caption{Measured spectrum of $^{90}$Sr electrons in the sample 9088-L-M at 200 V and $T=20^\circ$C. A fit of Gaussian-Landau convolution 
to the measured data is also shown.}
\label{fi:spectrum}
\end{figure}

An example of a recorded spectrum for sample 9088-L-M is shown in Fig. \ref{fi:spectrum}. The most probable charge was extracted from the fit of the convolution
of Gaussian and Landau function to the measured spectrum. 

It was observed in test beam measurements of R9088 devices that there is difference in charge collection between the metallized and non-metallized part of the detector of which the origin is not clear. The difference was few percent for non-irradiated, but could be as high as few tens percent close to the break down voltage in heavily irradiated sensors. In the presented measurements we could not separate between hits of metallized and non-metallized part, therefore the average is shown.

\subsection{Non-irradiated sensors}

\begin{figure}[!hbt]
\begin{tabular}{cc}
\epsfig{file=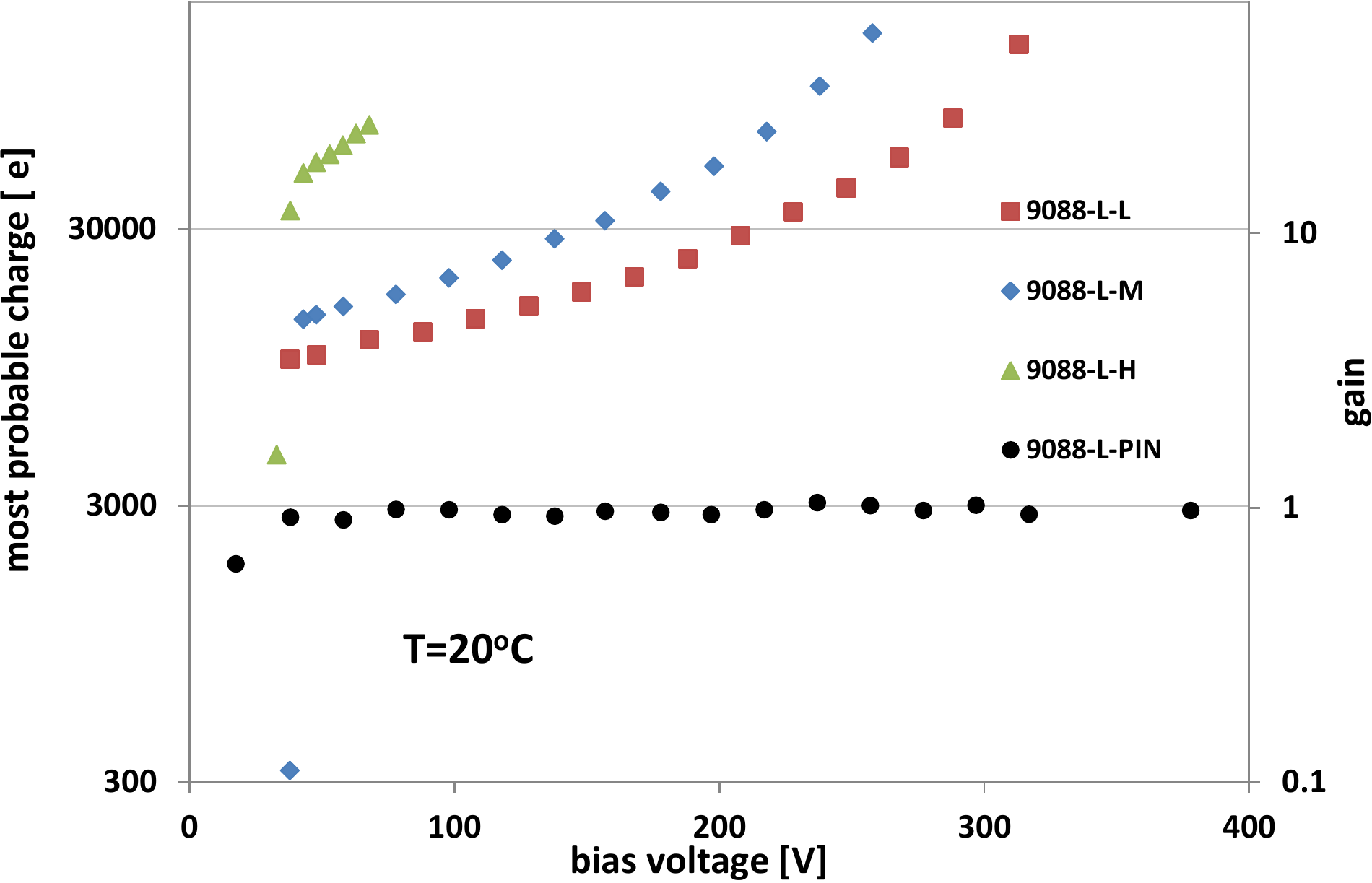,width=0.51\linewidth,clip=} & \epsfig{file=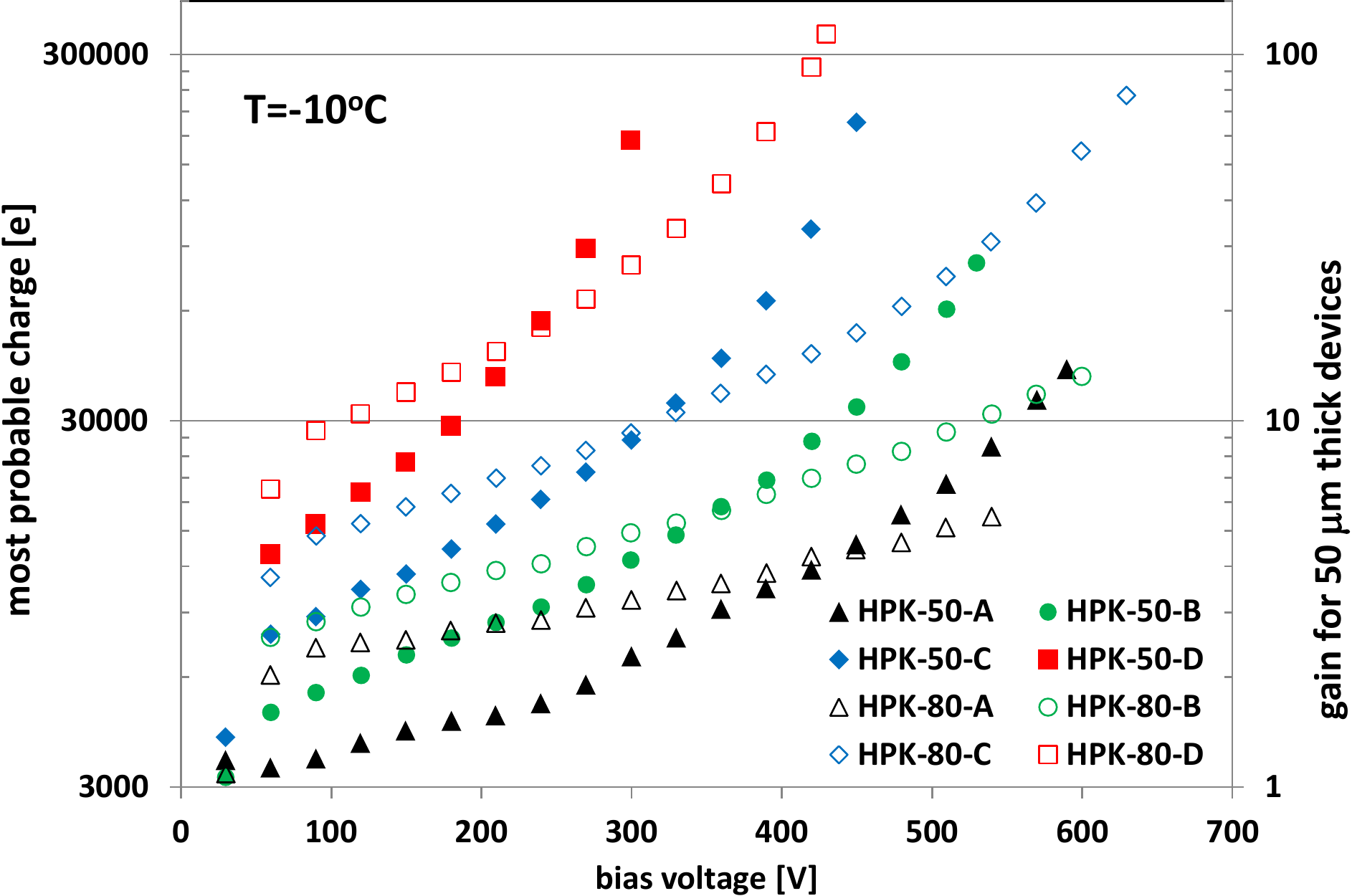,width=0.49\linewidth,clip=} \\
(a) & (b)
\end{tabular}
\begin{center}
\epsfig{file=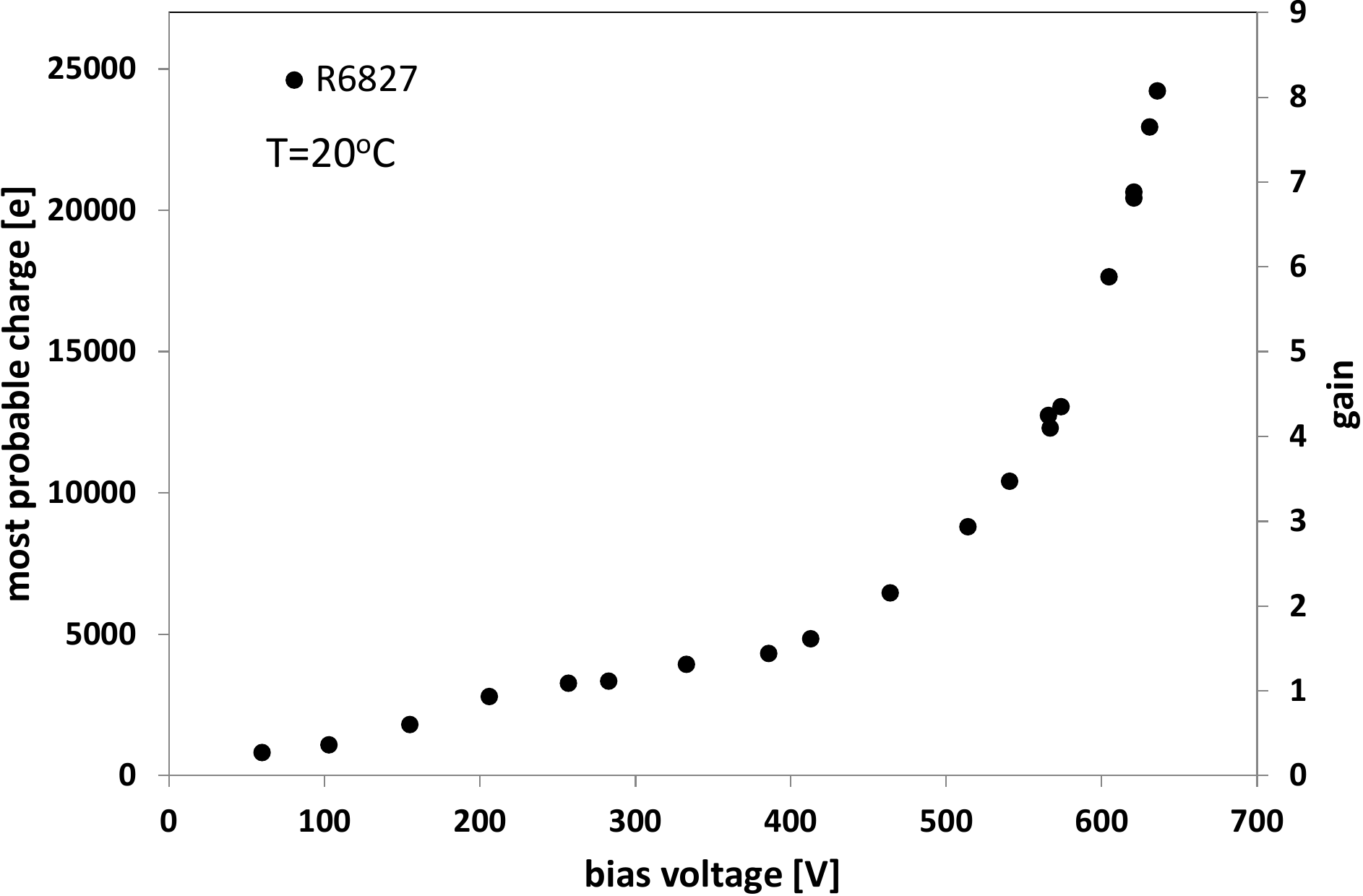,width=0.5\linewidth,clip=} \\
(c)
\end{center}
\caption{Dependence of measured most probable charge on bias voltage for different non-irradiated detectors: (a) R9088  (b) HPK and (c) R6827. The gain scale for HPK device holds for 50 $\mu$m thick devices only. Note the logarithmic scale in (a) and (b). }
\label{fi:QNirr}
\end{figure}

The measured signal for different investigated detectors is shown in Fig. \ref{fi:QNirr}. The gain of the devices 
is here defined as the most probable charge divided with the signal in a standard diode (PIN) of the same 
thickness $M=Q_{LGAD}/Q_{PIN}$ ($Q_{PIN}=2870$ e for R9088). 
Gains of up to almost one hundred were achieved. A clear dependence of most probable charge on different gain layer doping concentrations 
can be observed. A remarkable difference given the only few percent difference between implantation doses. The control of the gain and breakdown 
performance requires therefore excellent process control. The benefit of thicker detectors of more generated primary charge
can be (over) compensated with higher bias voltages due to average electric fields achieved leading to higher gains (see Fig. \ref{fi:QNirr}b). 
Thicker detectors also have smaller capacitance, hence noise, but the contribution due to 
Landau fluctuations to timing resolution is worse \cite{LGAD-Thin}. In addition the induced current amplitude 
is smaller at the same gain in thicker detectors  ($I\sim M/d$). Hence thicker detectors are suitable when smaller 
noise or limited voltage is required. The benefits of large breakdown voltages can be best seen in Fig. \ref{fi:QNirr}c. The R6827 device 
exhibits amplification only after substantial over-depletion, required to reach fields high enough for impact ionization.

\begin{figure}[!hbt]
\begin{center}
\epsfig{file=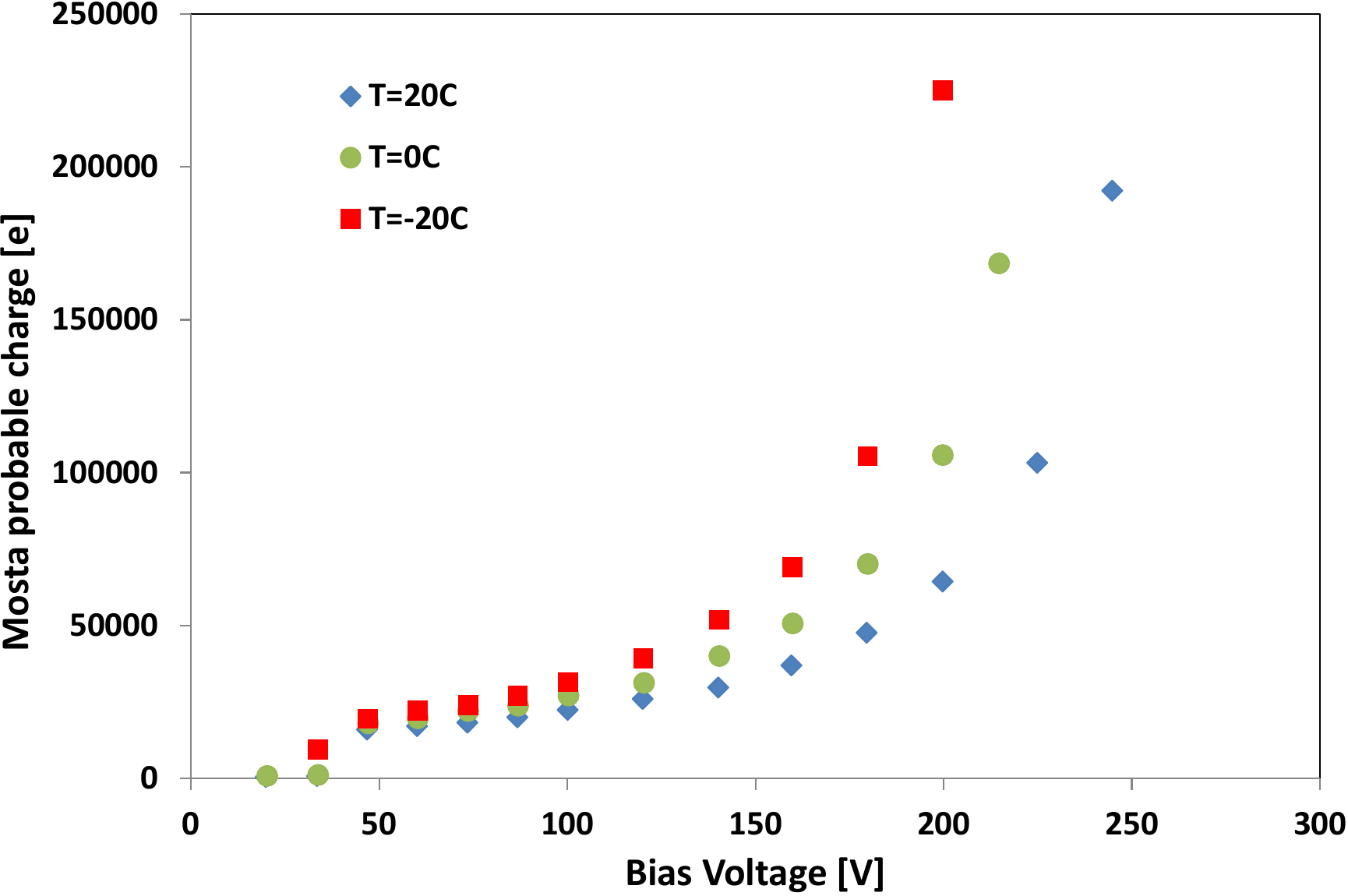,width=0.5\linewidth,clip=}
\end{center}
\caption{Dependence of collected charge in 9088-L-M on voltage at different temperatures. The highest votlage applied was close to the break-down votlage.}
\label{fi:QNirrTdep}
\end{figure}

The gain dependence on temperature was investigated for R9088 medium dose devices and is shown in Fig. \ref{fi:QNirrTdep}. 
A strong temperature dependence can be seen at high bias voltages (gains). The impact ionization coefficients become 
strongly temperature dependent at high fields \cite{II}. Around 30\% increase in gain can be assumed at $M \approx 10$ between $T=-20^\circ$C 
and $T=20^\circ$C. Influence of temperature on gain is therefore more significant in thin sensors where higher fields can be reached. 
Smaller gain at higher temperatures can be offset by applying higher bias as the break down voltage also increases with temperature. 

\subsection{Neutron irradiated sensors}

\begin{figure}[!hbt]
\begin{tabular}{cc}
\epsfig{file=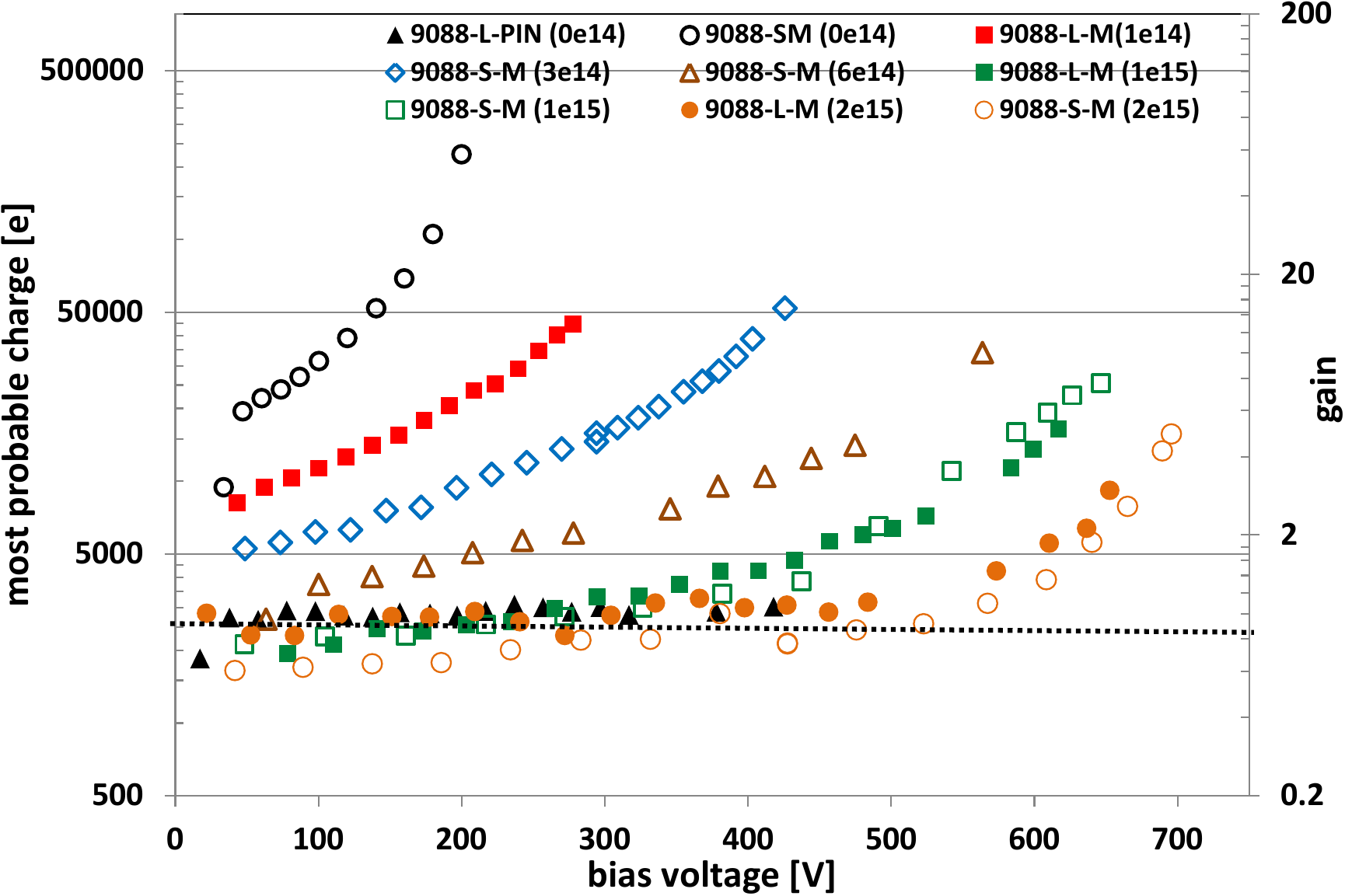,width=0.5\linewidth,clip=} & \epsfig{file=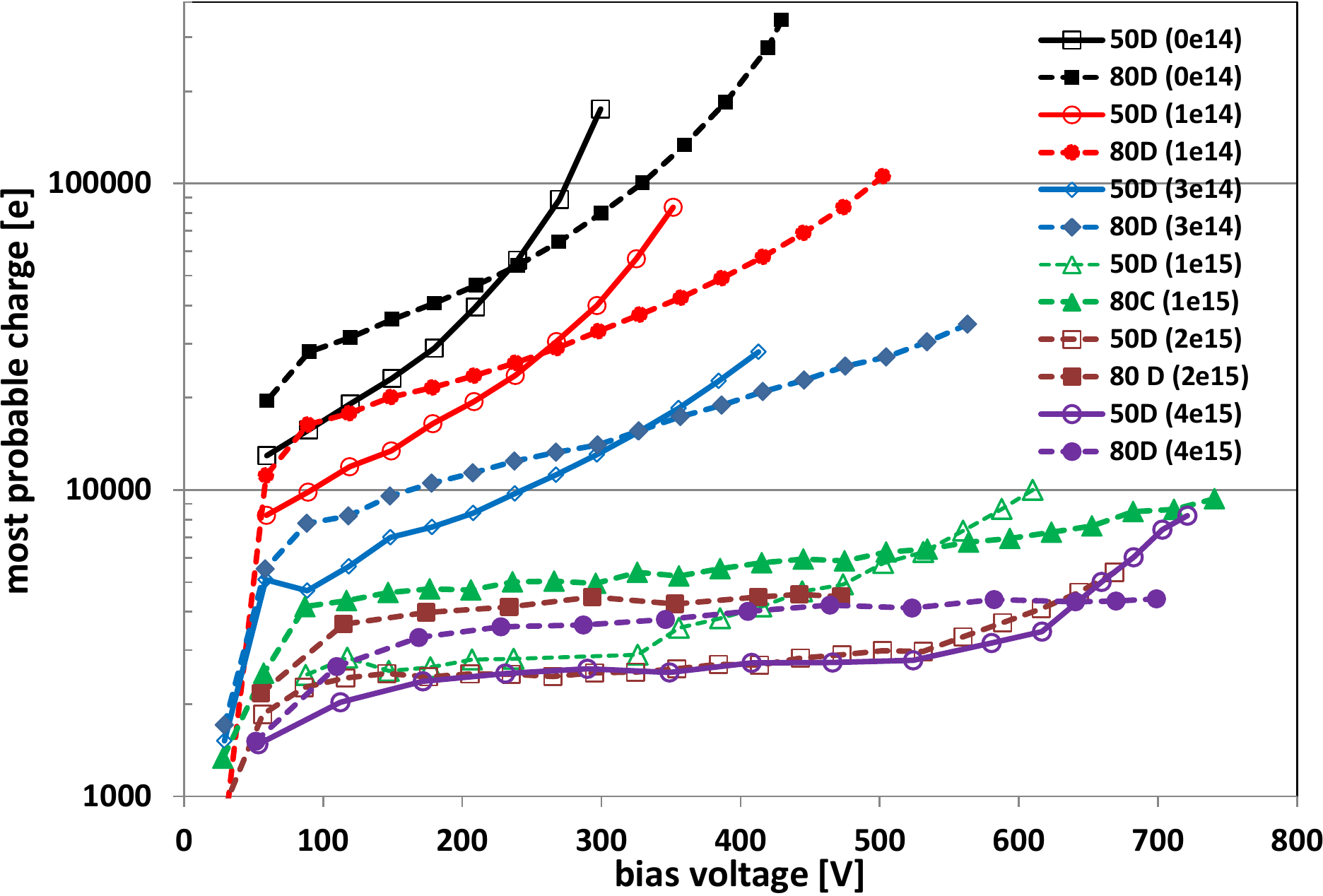,width=0.5\linewidth,clip=} \\
(a) & (b) 
\end{tabular}
\begin{center}
\epsfig{file=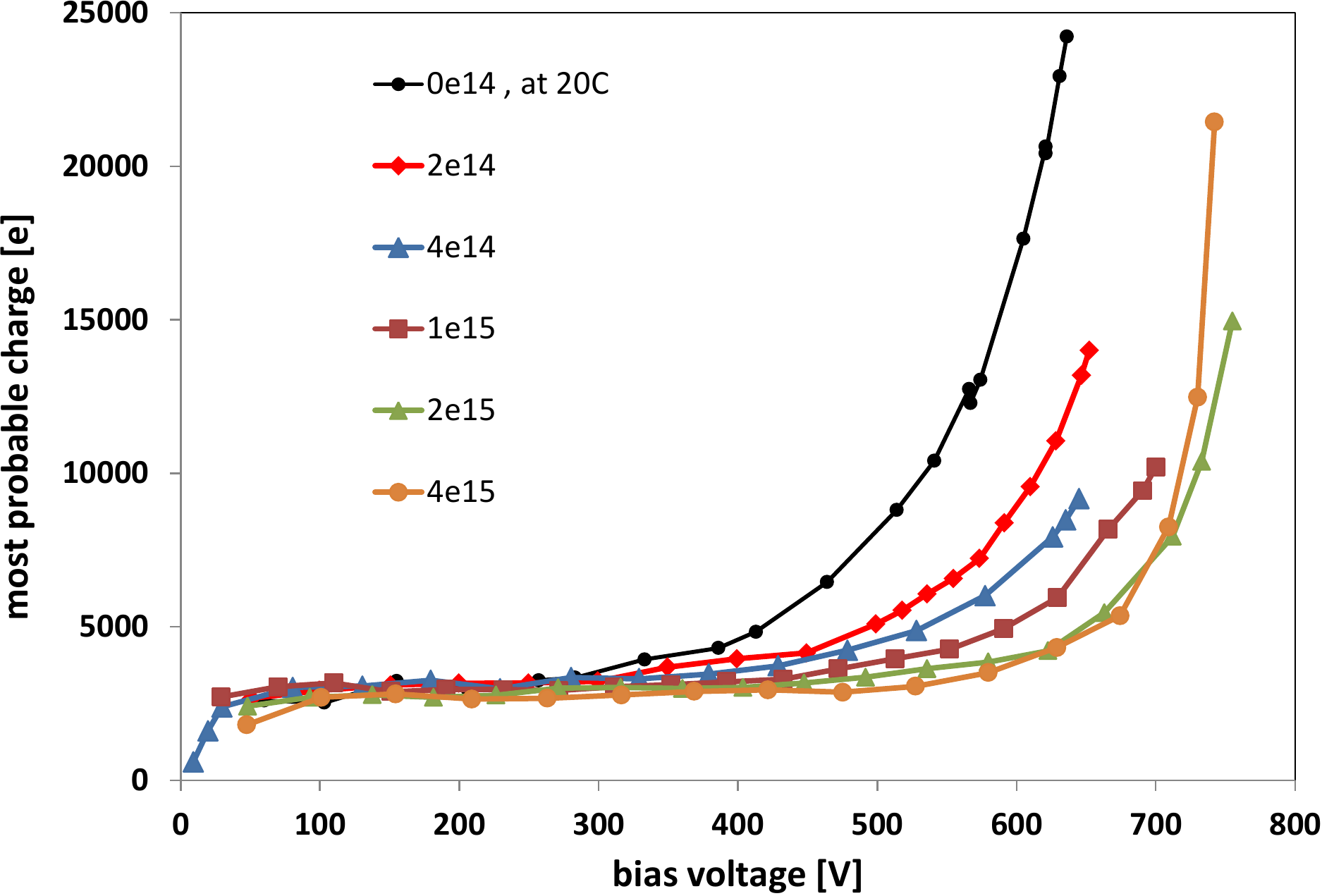,width=0.5\linewidth,clip=}  \\
(c)
\end{center}
\caption{Dependence of most probable charge on voltage at different fluences for irradiated devices from: (a) R9088 (b) HPK and (c) R6827. 
The measurements were done at $T=-10^\circ$C. The equivalent fluences are given in [cm$^{-2}$]. The dashed line in (a) denotes $Q_{PIN,nirr}$.}
\label{fi:Qirr}
\end{figure}

The collected charge in irradiated sensors is shown in Fig. \ref{fi:Qirr}. The definition of the gain requires the knowledge of 
$Q_{pin}$ for fully depleted irradiated sensor and is often replaced by $Q_{PIN,nirr}$ for easier calculation of the charge. Only after 
trapping becomes substantial these two definitions differ, which for thin detectors occurs at fluences larger than few $10^{15}$ cm$^{-2}$.

A large decrease of charge with fluence is evident. 
At lower fluences the gain is still substantial after full depletion of the device ($<100$ V) while for high fluences 
the gain remains visible only at highest voltages. 
The decrease of gain is attributed to initial acceptor removal in the multiplication layer \cite{LGAD-radhard} and will be addressed in the next section. 
The gain loss at lower fluences can be compensated by applying higher bias voltage (see Fig. \ref{fi:EFieldSchematic}), but a full compensation is not possible 
anymore at fluences $\Phi_{eq}>10^{15}$ cm$^{-2}$.

Electric fields at highest bias voltages 600-700 V (on average 12-15 V/$\mu$m) are sufficient to cause multiplication in the large 
part of the bulk and $\sim$50 $\mu$m devices break down in a very narrow voltage interval around 700 V at high fluences. Although the 
break-even voltage in collected charge for 50 $\mu$m and 80 $\mu$m devices (Fig. \ref{fi:Qirr}b) shifts to ever larger values due 
to reduced gain, thin devices nevertheless always outperform thick ones in the investigated fluence range.

The residual concentration of initial acceptors in the gain layer plays a larger role in thin than in thick detectors after irradiation. 
Although not sufficient for yielding gain immediately after depletion ($Q$ is constant over a large voltage interval) they lead 
to improved gain at high voltages. This effect can be observed in all devices and it is best illustrated in Fig. \ref{fi:Qirr}c, 
for the devices from R6827 where the initial doping was low enough to observe this behavior already before irradiation. The 
increase of  voltage required for the onset of multiplication with fluence can be clearly observed. At the highest two 
fluences ($\Phi_{eq} \geq 2 \cdot 10^{15}$ cm$^{-2}$) the collected 
charge dependence on voltage is almost the same. Larger concentration of deep acceptors in the bulk may even lead to larger 
gains at voltages close to break down voltage.

\begin{figure}[!hbt]
\begin{tabular}{cc}
\epsfig{file=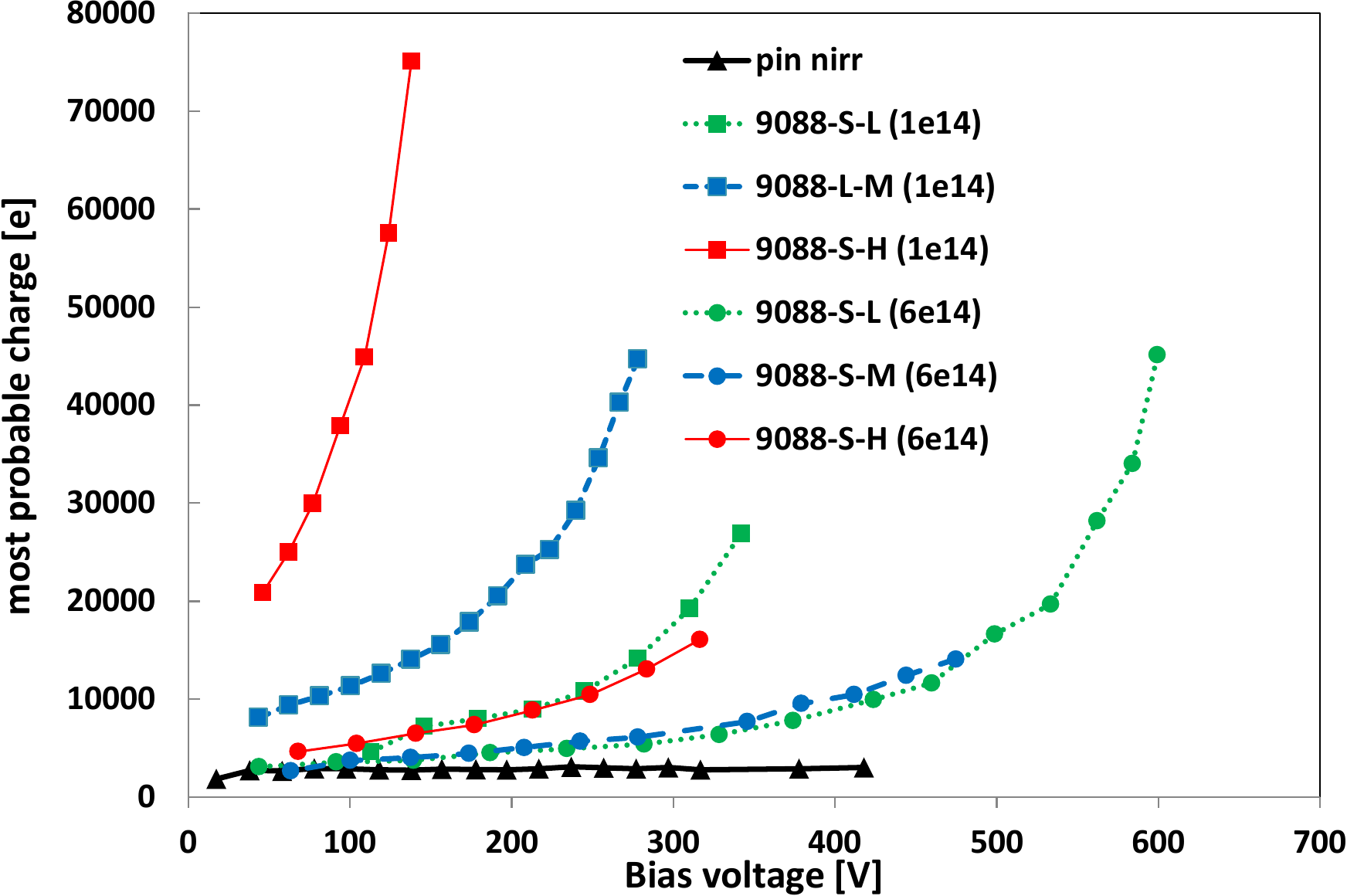,width=0.5\linewidth,clip=} & \epsfig{file=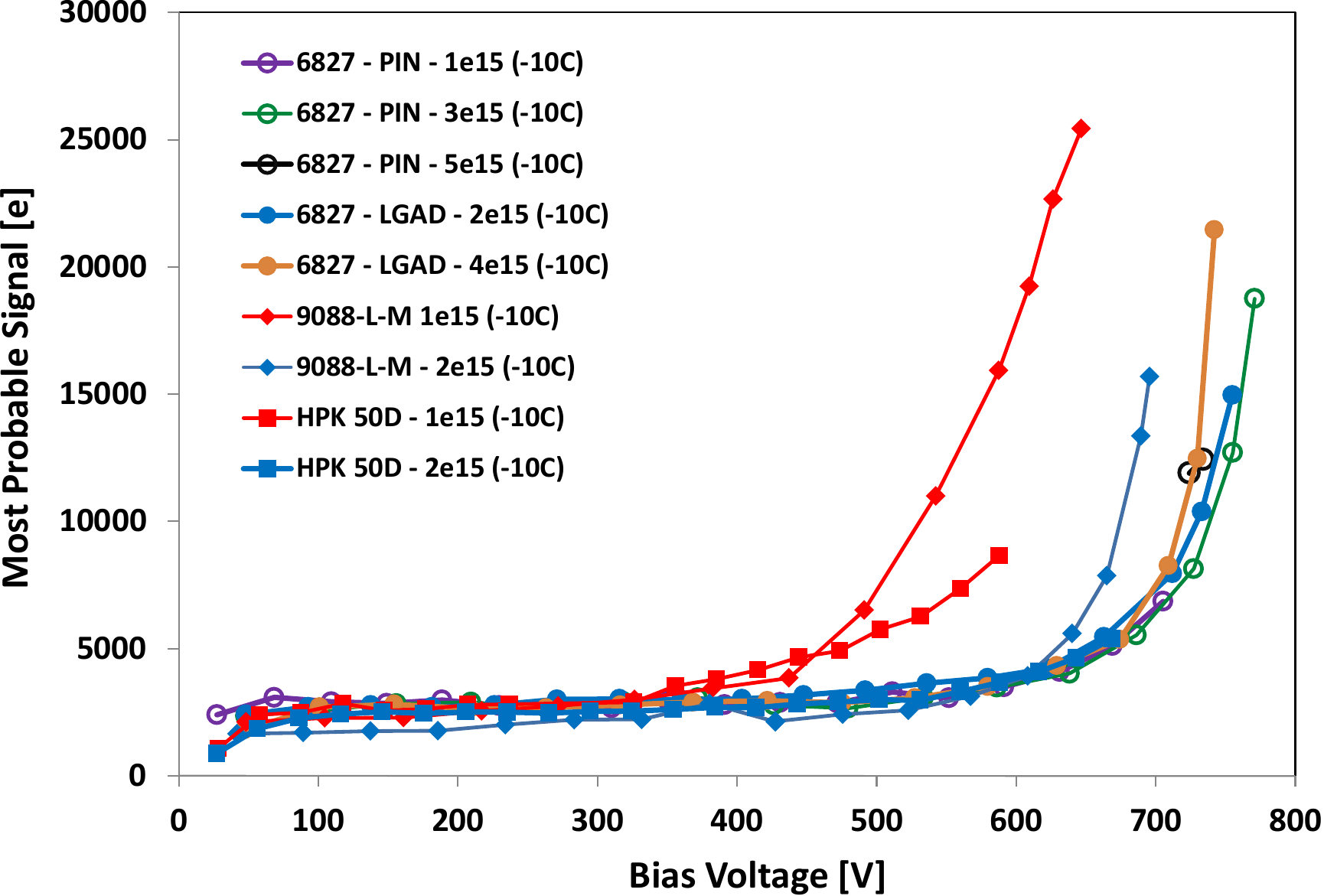,width=0.5\linewidth,clip=} \\
(a) & (b) 
\end{tabular}
\caption{(a) Dependence of most probable charge on voltage: (a) for irradiated R9088 devices with different initial doping 
concentration, (b) for devices from different sets irradiated to fluences $\Phi_{eq} \geq 10^{15}$ cm$^{-2}$ at $T=-10^\circ$C. 
The equivalent fluences are in [cm$^{-2}$]. }
\label{fi:DoseEffect}
\end{figure}

The removal of initial acceptors favors the choice of highest possible doping of the multiplication layer. This is shown 
in Fig. \ref{fi:DoseEffect}a, where the devices from R9088 are compared. However, although the difference still remains after 
$\Phi_{eq}=6\cdot 10^{14}$ cm$^{-2}$, it is becoming smaller. The same observation holds also for HPK samples. At fluences 
above $\Phi_{eq} \geq 2 \cdot 10^{15}$ cm$^{-2}$ the performance of devices becomes universal regardless of producer or initial 
doping, with even the same performance of the LGAD device and diodes without multiplication layer (see Fig. \ref{fi:DoseEffect}b). Moreover, 
there is also little effect of irradiation fluence. The effect of trapping seems to be small and with precision of our
measurements is estimated to be less than 20\% difference in most probable charge between non-irradiated sensors and 
those to $\Phi_{eq}=4\cdot 10^{15}$ cm$^{-2}$ before the onset of multiplication.

\begin{figure}[!hbt]
\begin{center}
\epsfig{file=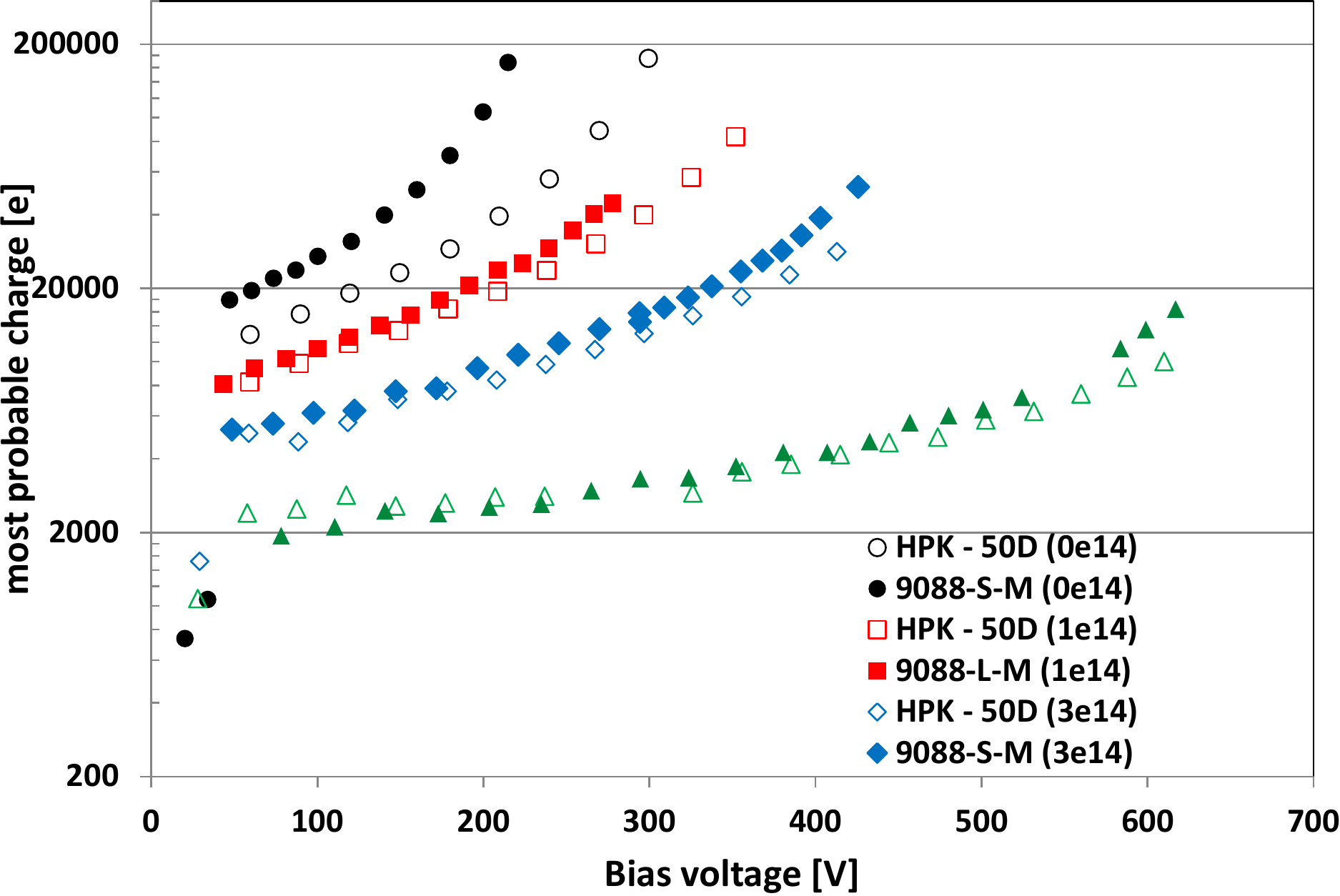,width=0.5\linewidth,clip=}
\end{center}
\caption{Comparison of most probable charge measured at $T=-10^\circ$C in CNM and HPK sensors of similar initial gain after irradiation.}
\label{fi:HPK-CNM}
\end{figure}

The rate of initial acceptor removal depends on concentration \cite{AR}, therefore different doping profiles could lead to different 
post irradiation gains, even when similar before irradiation. The two producers have different processes, but the comparison of 
devices with similar initial gain exhibited little difference in charge vs. voltage dependence as shown 
in Fig. \ref{fi:HPK-CNM}. Whether this is a coincidence or the parameter space in device processing is so restricted 
that different producers converge to similar doping profile shapes remains an open question. 

\begin{figure}[!hbt]
\begin{tabular}{cc}
\epsfig{file=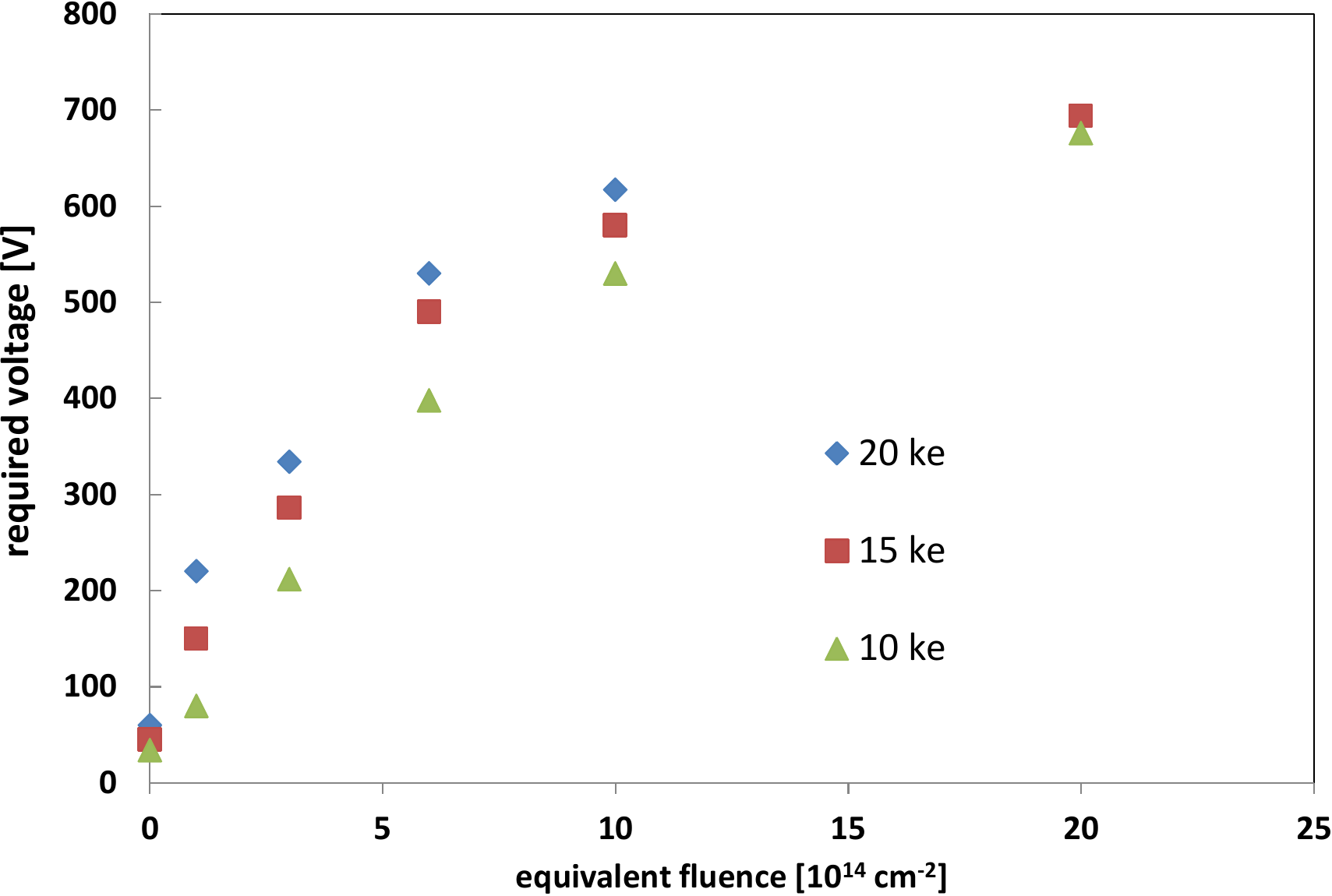,width=0.5\linewidth,clip=} & \epsfig{file=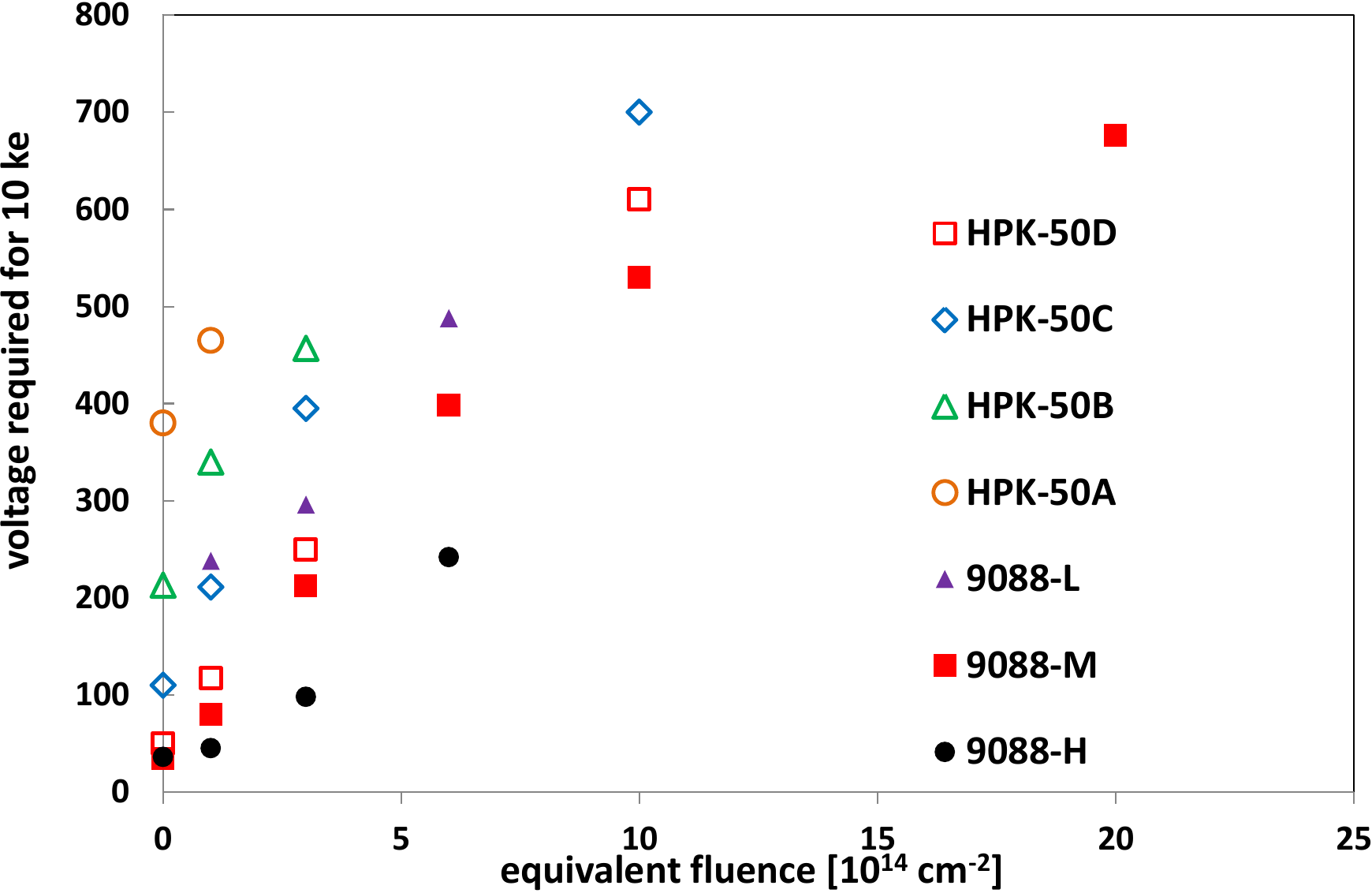,width=0.5\linewidth,clip=} \\
(a) & (b)
\end{tabular}
\caption{(a) Dependence of voltage required for the collection of given most probable charge in R9088-S-M devices. (b) Voltage required for the collection of 10 ke for different sample sets. Measurements were done at $T=-10^\circ$C.}
\label{fi:V10keV}
\end{figure}

The collected charge determines to a large extent the timing resolution of the sensors once the applied voltage is sufficient to saturate the drift velocity 
in the entire bulk. It is assumed that shot noise and by that jitter can be kept under control, due to small foreseen integration times of $\sim$1 ns. 
Saturation of drift velocities requires electric fields in the bulk of few V/$\mu$m. This is almost always achievable already after low fluences or for devices with lower initial 
gain layer doping, but not necessarily for high gain layer doping which can lead to early break down (see Fig. \ref{fi:QNirr}). The bias voltage therefore 
determines the gain and by that to a large extent the timing resolution of the devices. The figure of merit for a detector is therefore voltage 
required for given charge. The further the voltage is from the breakdown voltage the safer is the operation of the detector. 
In Fig. \ref{fi:V10keV}a the dependence of the required voltage for the collection of 10,15 and 20 ke 
is shown for R9088 devices as a function of equivalent fluence. The difference in voltage required for 10 and 20 ke is around 150 V and decreases with fluence as the charge rise becomes steeper when approaching the breakdown voltage (see Fig. \ref{fi:Qirr}). Timing resolution of around 50 ps corresponds roughly to 20 ke \cite{JL,HS}. The benefit of having higher initial doping concentration is shown in Fig. \ref{fi:V10keV}b for different detectors from sets R9088 and HPK. Therefore higher initial doping will keep the benefits of LGADs to larger fluences and would allow to operate detectors at lower voltages.

The three sets of samples were produced on different substrates, which seem not to influence the charge collection performance.

\subsection{Pion irradiated CNM sensors}

\begin{figure}[!hbt]
\begin{tabular} {cc}
\epsfig{file=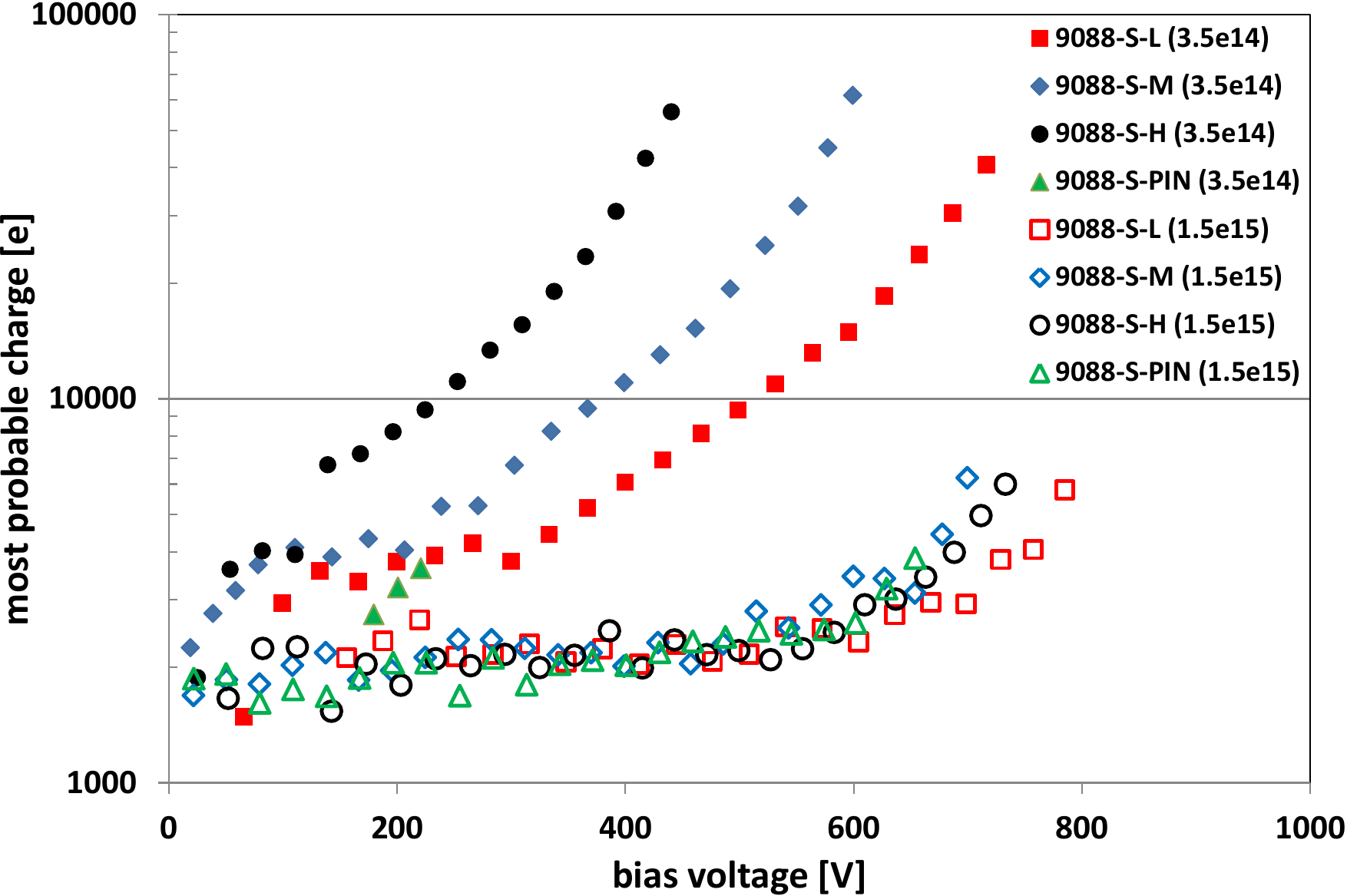,width=0.5\linewidth,clip=} & \epsfig{file=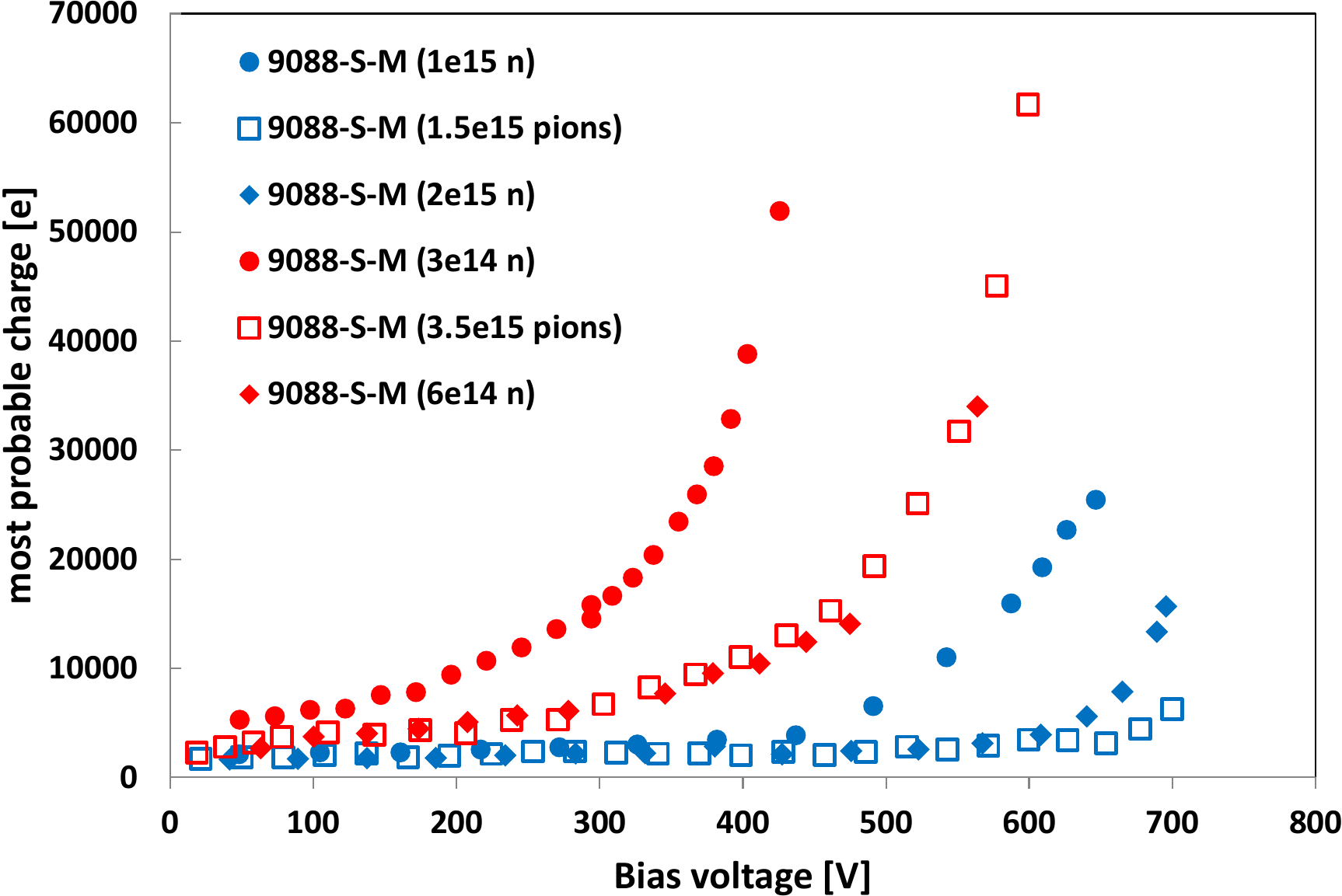,width=0.5\linewidth,clip=} \\
(a) & (b) 
\end{tabular}
\caption{(a) Dependence of collected charge on bias voltage for different pion irradiated samples at $T=-10^\circ$C. (b) Comparison charge collection of 9088-S-M samples irradiated with neutrons and pions.}
\label{fi:pions}
\end{figure}

At HL-LHC radiation damage will be caused by both charged hadrons and neutrons. It is long known that at the same non-ionizing energy loss (NIEL) 
the damage inflicted to the detector can be different regarding the irradiation particle type (NIEL hypothesis violation) \cite{RD48}. As most 
of the studies with irradiated LGADs were so far performed with neutrons it is important to check the acceptor removal and related loss of 
gain also after charged hadron irradiations. R9088 samples of all doping concentrations were irradiated with pions to equivalent fluences of 
$\Phi_{eq}=3.5\cdot 10^{14}$ cm$^{-2}$ and $\Phi_{eq}=1.55\cdot 10^{15}$ cm$^{-2}$. Measured charge vs. bias voltage for 
pion irradiated devices is shown in Fig. \ref{fi:pions}a. It seems that for low fluences the bulk depletes first and only then the gain layer which leads to a steep rise in collected charge. 
As for neutron irradiation, a higher gain is preserved for higher initial doping at medium fluence, whereas at high fluence the difference becomes small.

The required voltage for a given charge is larger for pion compared to neutron irradiated samples at comparable fluences (see Fig. \ref{fi:pions}b). 
The $Q-V$ curve for pion irradiated sample to $\Phi_{eq}=3.5\cdot 10^{14}$ cm$^{-2}$  
is similar to that of the neutron irradiated one at almost double equivalent fluence ($\Phi_{eq}=6\cdot 10^{14}$ cm$^{-2}$). This is true also 
for samples irradiated to high fluence.

It is therefore essential to more precisely evaluate the acceptor removal after charged hadron irradiations and also mixed irradiations (charged hadron and neutron irradiations) in order to be able to make predictions of detector operation at different locations in HL-LHC experiments.

\section{Leakage current}

\begin{figure}[!hbt]
\begin{tabular}{cc}
\epsfig{file=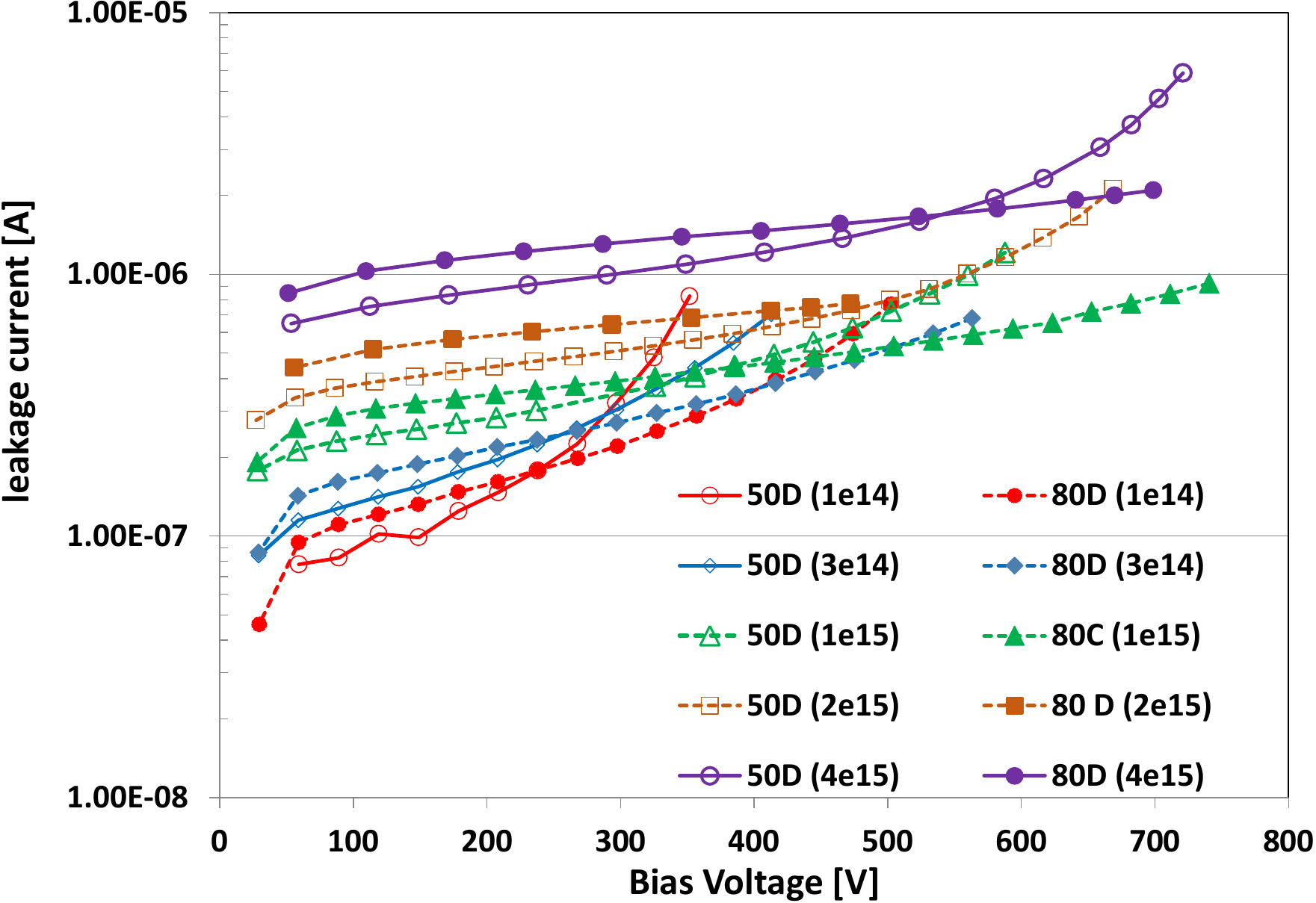,width=0.5\linewidth,clip=} &  \epsfig{file=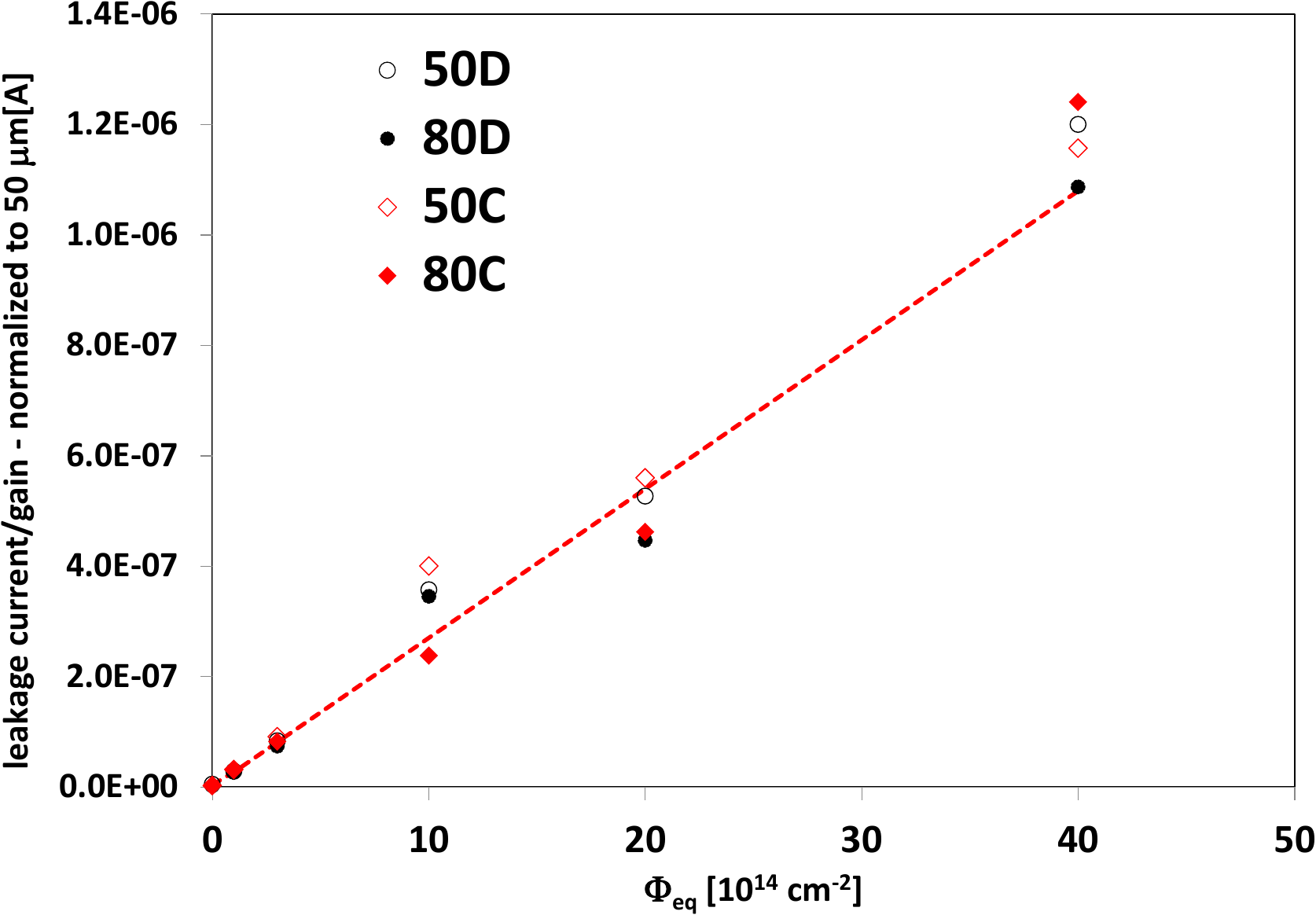,width=0.5\linewidth,clip=} \\
(a) & (b)
\end{tabular}
\begin{center}
\epsfig{file=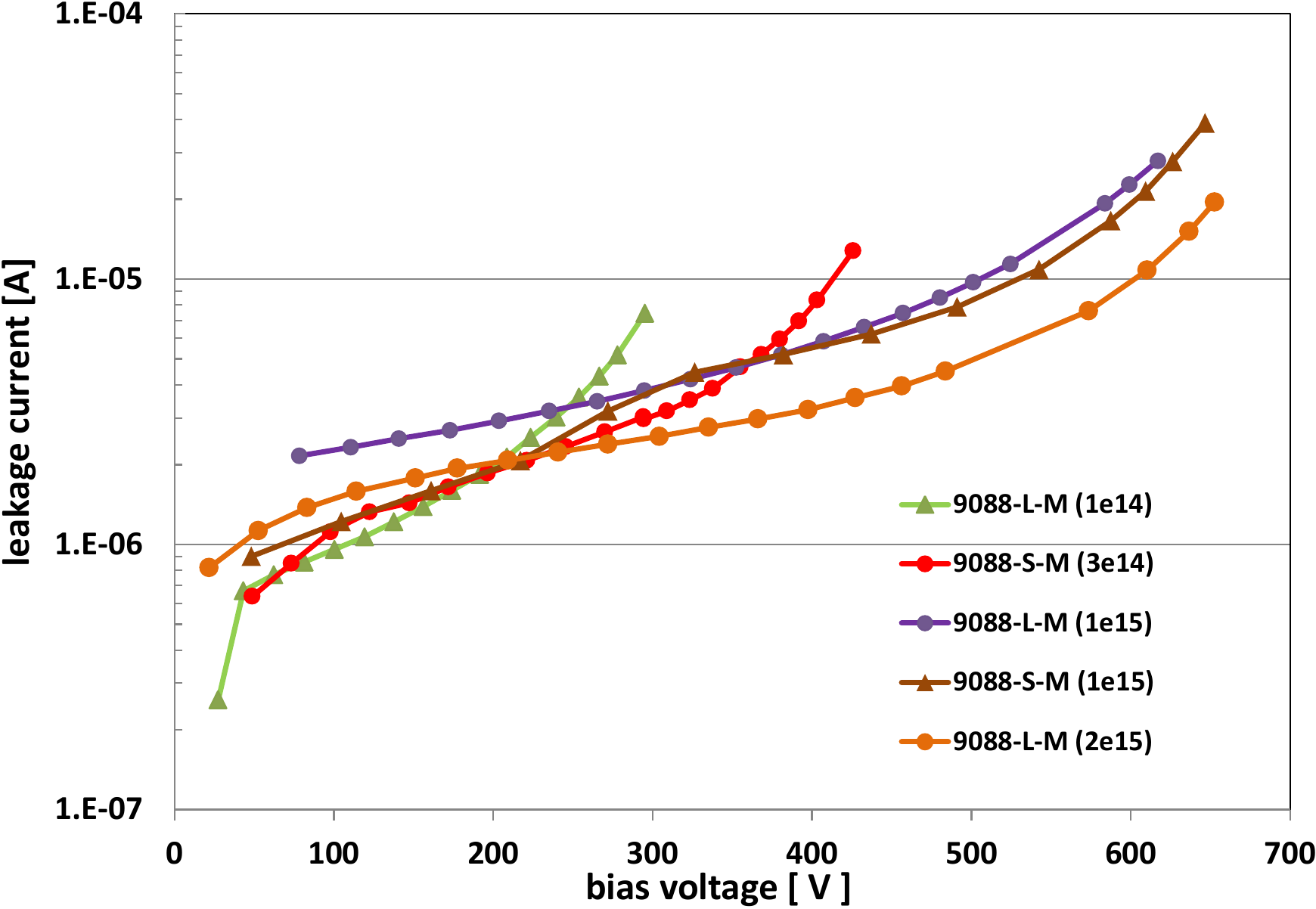,width=0.5\linewidth,clip=}  \\
\end{center}
\caption{(a) Leakage current for HPK samples measured at different neutron fluences given in [cm$^{-2}$]. (b) Leakage current for different HPK detectors after accounting for gain and thickness. The red dashed line denotes predicted/calculated generation current. (c) Same as (a) for R9088 samples. All measurements are shown at  $T=-10^\circ$C.}
\label{fi:Ileak}
\end{figure}

As has been shown in Ref. \cite{LGAD-radhard} the leakage current in LGAD devices follows 
\begin{equation}
I=M_I \,\, I_{gen}=M_{I} (V_{bias},\Phi_{eq})\,\, \cdot \alpha \,\cdot \, \Phi_{eq}\, \cdot \, d\, \cdot\, S \quad,
\label{eq:Ileak}
\end{equation}
where $I_{gen}$ is the generation current, $M_I$ the current multiplication factor, $\alpha(-10^\circ$C$)=2.14\cdot10^{-18}$ A/cm the leakage current damage constant \cite{RD48}, $d$ the detector thickness and $S$ the active surface. 
For some samples from CNM runs the leakage current was un-usually high before irradiation which often led to a discrepancy between the calculated
and measured currents \cite{LGAD-radhard}. The leakage current dependence on fluence is not linear as the decrease of $M_I$ compensates the increase of 
the generation current. The measured dependence of the leakage current for R9088 and HPK devices is shown in Fig. \ref{fi:Ileak}. With $M_I$ assumed to be 
the same as charge gain $M$ determined from charge collection measurements, the agreement between calculated and measured current for HPK sensors is 
within 20\% (mainly from uncertainty in temperature). An average of the current measurements above the full depletion voltage divided by the gain and normalized to $50\,\,\mu$m thickness is shown in Fig. \ref{fi:Ileak}b. Such a good agreement between measured and calculated currents allows for an estimate of the gain from current measurements. The measured 
current in R9088-L-M samples is given in Fig. \ref{fi:Ileak}c and similar conclusion as for HPK sensors holds. Note that the larger current for R9088 devices
 than for HPK is a consequence of the different active area. 

\section{TCT studies of multiplication layer}

\begin{figure}[!hbt]
\begin{tabular}{cc}
\epsfig{file=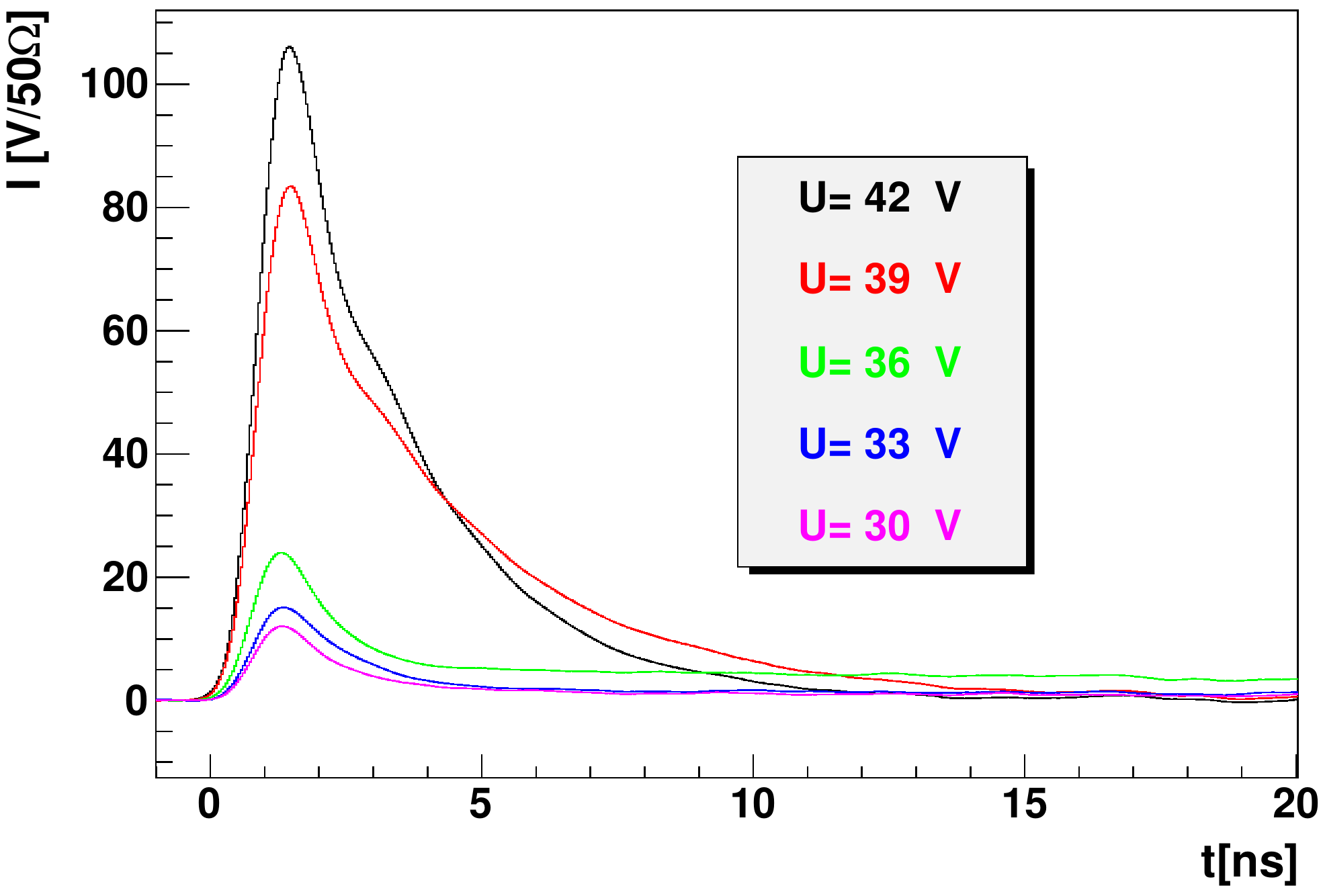,width=0.5\linewidth,clip=} & \epsfig{file=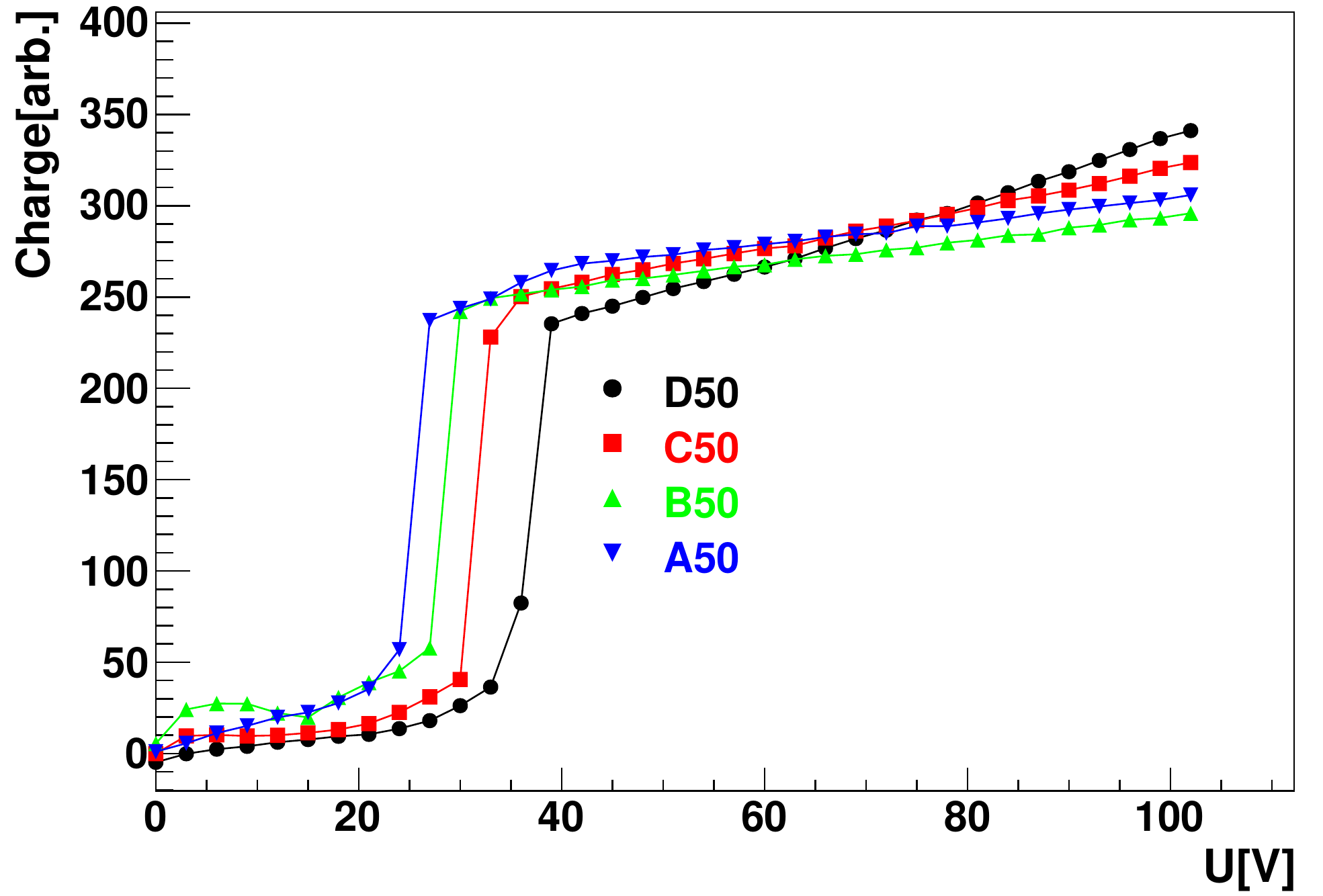,width=0.5\linewidth,clip=}  \\
(a) & (b) \\
\epsfig{file=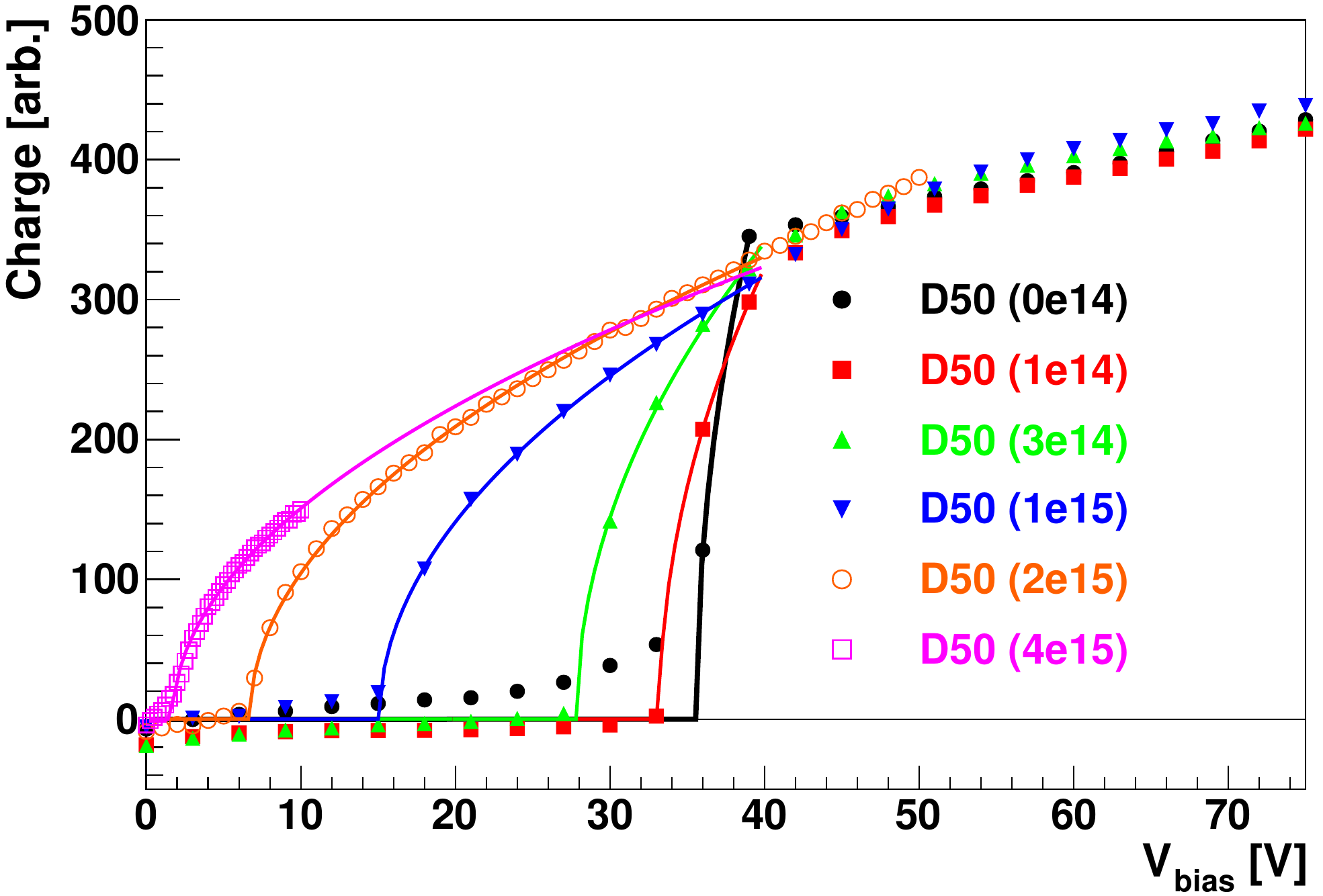,width=0.5\linewidth,clip=} & \epsfig{file=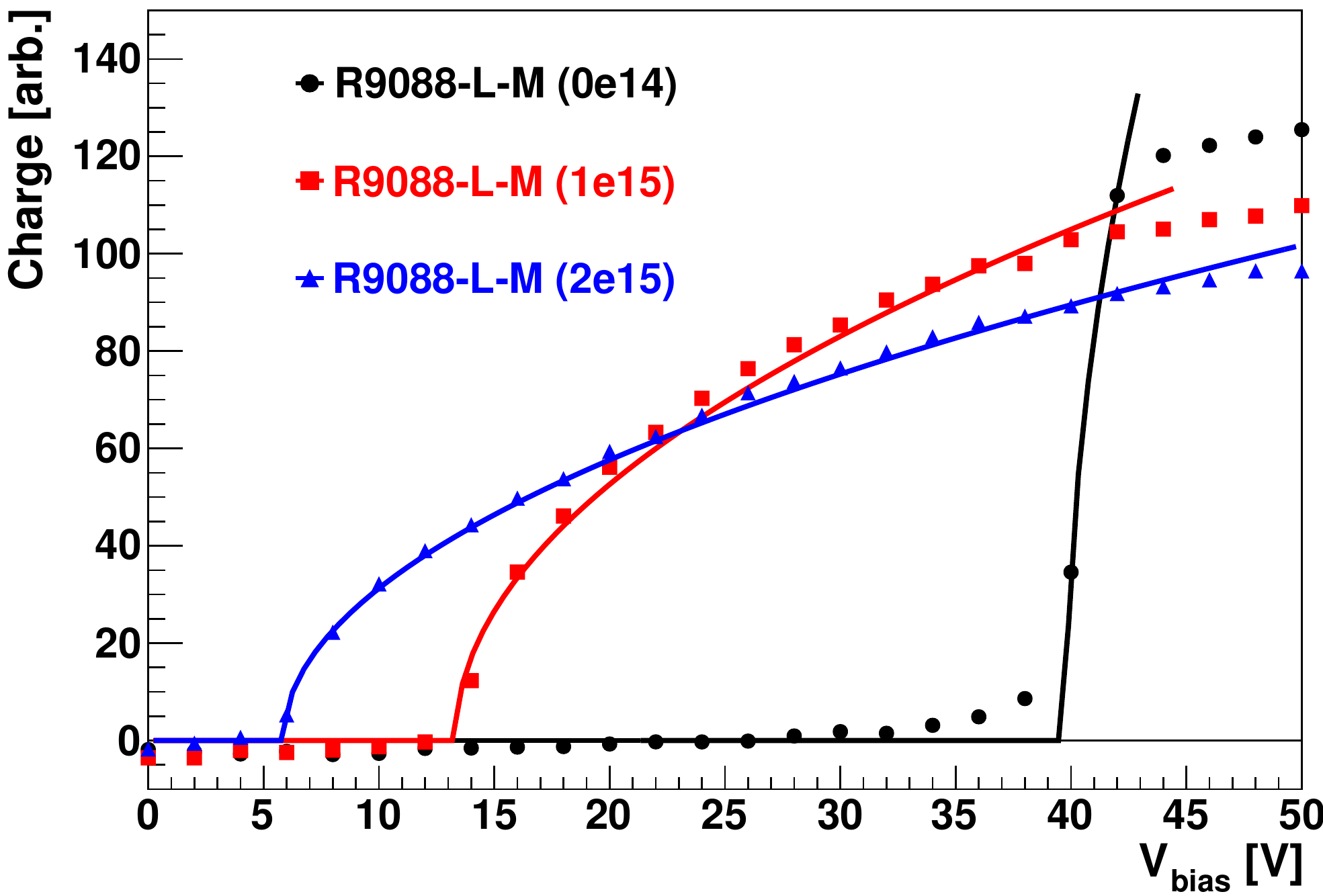,width=0.5\linewidth,clip=} \\
(c) & (d) \\
\end{tabular}
\caption{(a) Induced currents in a HPK-50D device. (b) Induced charge for different non-irradiated HPK-50 devices. 
(c) Induced charge for HPK-50D devices irradiated to different fluences. (d) Induced charge for R9088 devices irradiated to different fluences. The fit of $\sqrt{V_{bias}-V_{mr}}$ to the data is also shown. All measurements were done at $T=-10^\circ$C. }
\label{fi:TCTSignalsFoot}
\end{figure}

The transient current technique (TCT) is an ideal tool to study the properties of the gain layer. The voltage required to deplete the p$^+$ 
layer can be probed by observing the induced currents after front illumination of the detector by light of 
short penetration depth ($\sim$3 $\mu$m at $\lambda=660$ nm; pulse width FWHM$\sim$400 ps, 500 Hz repetition). Once the p$^+$ layer is 
depleted the induced current increases as more carriers drift.  An example of induced current pulses
for a non-irradiated HPK detector is shown in Fig. \ref{fi:TCTSignalsFoot}a, where the steep increase of the current at 39 V indicates
the depletion of the multiplication layer. The dependence of charge (current integral in 20 ns) on bias voltage is shown in Fig. \ref{fi:TCTSignalsFoot}b.
The difference in p$^+$-layer depletion voltages for samples with different implantation doses is clearly visible. For irradiated sensors the bulk becomes highly resistive and the 
electrode (ground) is not effectively at the border of the depleted region anymore. Therefore
the current is induced only after the carriers drift over a significant distance (weighting potential)  without recombination. 
For constant space charge the depleted region in the p bulk grows as $\sqrt{V_{bias}}$, and by that also the induced charge.

\begin{figure}[!hbt]
\begin{tabular}{cc}
\epsfig{file=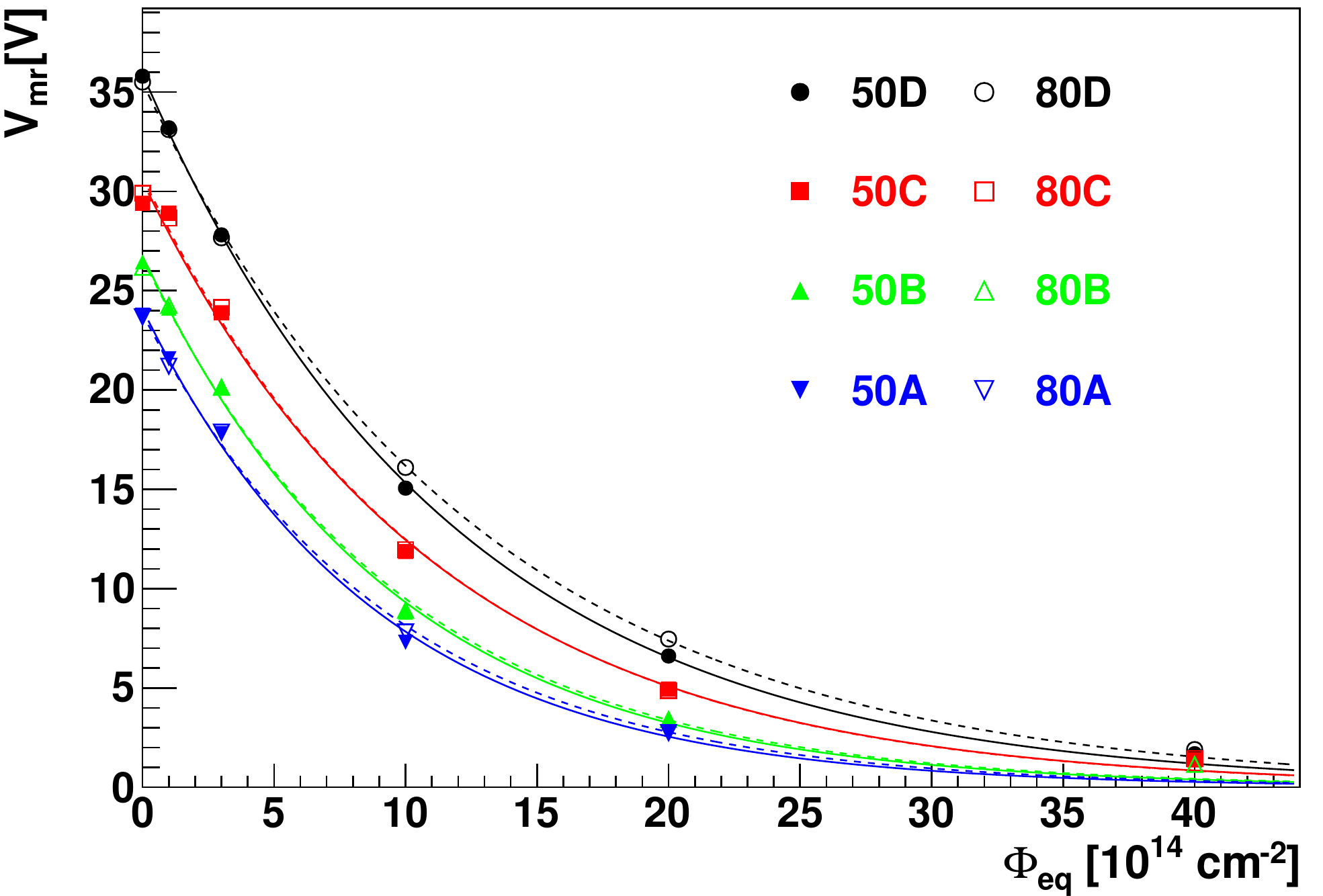,width=0.5\linewidth,clip=} & \epsfig{file=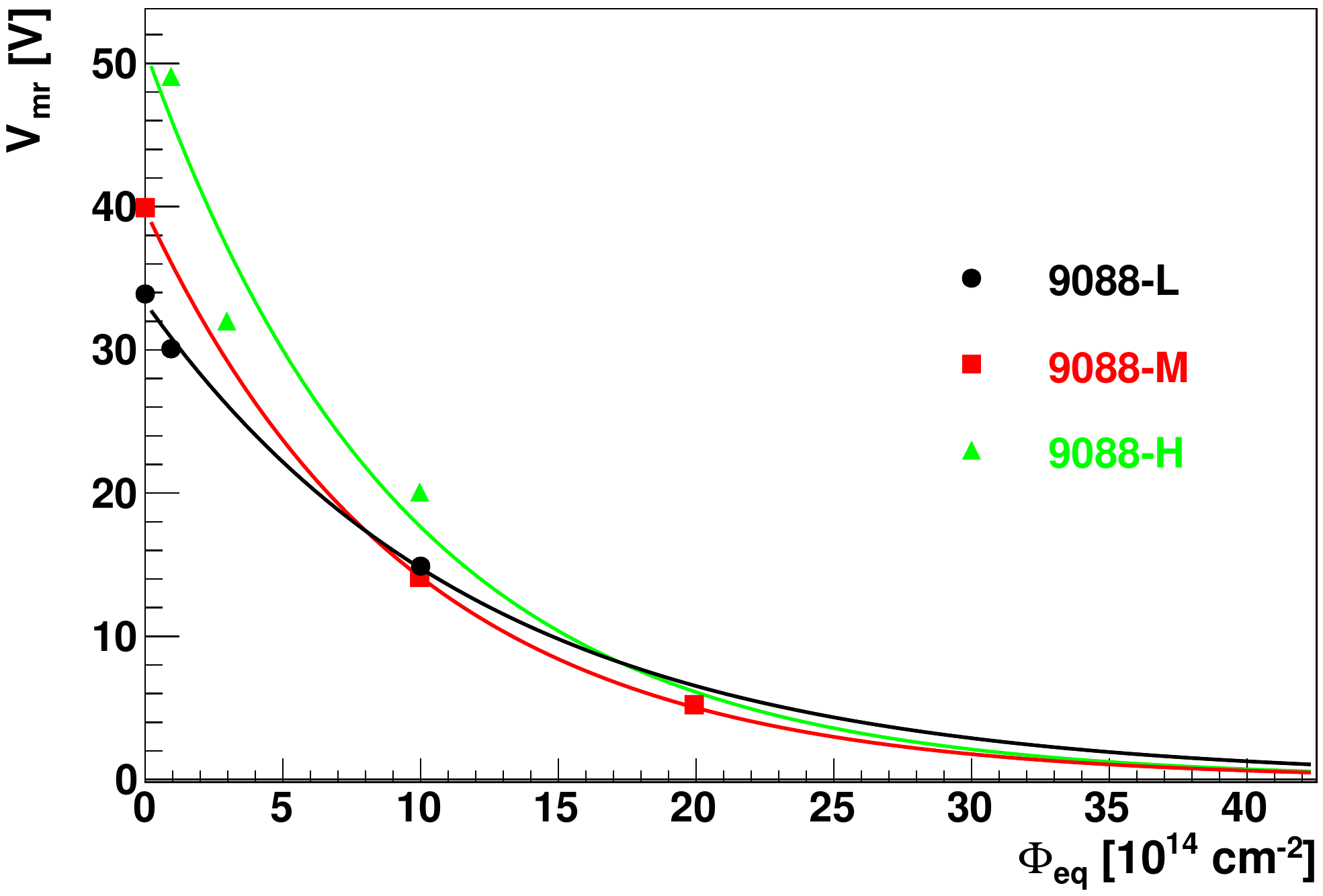,width=0.5\linewidth,clip=}  \\
(a) & (b) \\
\end{tabular}
\caption{Evolution of the multiplication layer depletion voltage $V_{mr}$ with neutron fluence 
for: (a) HPK devices and (b) CNM R9088 devices. The fit of Eq. 5.1 to the data is also shown.}
\label{fi:Vmr}
\end{figure}

The voltages required for the depletion of the multiplication 
layer $V_{mr}$ were obtained by fitting $Q \propto \sqrt{V_{bias}-V_{mr}}$ to the measured charge. 
As seen in Figs. \ref{fi:TCTSignalsFoot}c,d $V_{mr}$ decreases with equivalent fluence. 
If  the boron removal occurs with the same rate everywhere in the p$^+$ layer, 
then $V_{mr}$ is proportional to the average concentration of Boron $N_{B}$ and the 
evolution of $V_{mr}$ with fluence can be described as 
\begin{equation}
N_{B}=N_{B,0}\,\,\exp( -c \,\, \Phi_{eq}) \quad \Rightarrow \quad V_{mr} \approx V_{mr,0}\,\,\exp(-c \,\, \Phi_{eq}) \quad,
\label{eq:IAremoval}
\end{equation} 
where $c$ is the removal constant, 
$N_{B,0}$ the initial doping concentration and $V_{mr,0}$ the multiplication layer depletion voltage.
The extracted $V_{mr}$ at different fluences and the Eq. \ref{eq:IAremoval} fit to the data 
for all 
HPK devices are shown in Fig. \ref{fi:Vmr}a. The free parameters of the fit 
were $c$ and $V_{mr,0}$. It can be seen that the depletion of the $p^+$ layer does not depend on the device thickness. 
This confirms the assumption that the main mechanism of the  
effective acceptor removal is the deactivation of boron rather than the compensation of initial acceptors by deep donor levels. A larger 
concentration of thermally generated free carriers in the thicker detector would impact $V_{mr}$ in the latter case.
The obtained values were $c=10.9, 10.4, 9.0\,\, \mathrm{and}\,\, 8.5 \cdot10^{-16}$ cm$^2$ and $V_{mr,0}=23.9, 26.7, 30.6, 35.8$ V 
for A,B,C and D devices. 

The data for R9088 are shown in Fig. \ref{fi:Vmr}b. The analysis here is influenced by the fact that the depletion of samples with high 
leakage current exhibits a so called ``double junction'' effect which effectively prevents an accurate determination for some of the sensors. 
Similar values of $c=8.1, 10.3, 10.6 \cdot10^{-16}$ cm$^2$ and $V_{mr,0}=33.3, 39., 49$ V were obtained also 
for 9088-L,M and H devices. Given the difference of only around 12\% between the maximum and minimum implantation dose for HPK and R9088 devices 
the reasons for significantly larger difference in $V_{mr}$ are not clear to us.

The removal rates for both investigated sets are similar to those obtained previously in 300 $\mu$m thick devices produced by CNM \cite{LGAD-radhard}. 
Comparable removal rates of initial acceptors for both producers point on one side to similar doping profiles for both, but also to 
the independence of the removal rate on all other process influenced parameters.

\section{Conclusions}

Systematic charge collection measurements were performed on three sets of non-irradiated and irradiated thin, 
50 $\mu$m and 80 $\mu$m, LGAD detectors produced by CNM and HPK. The measurements showed a decrease of the collected 
charge/gain with fluence in agreement with initial acceptor removal in the multiplication layer. The removal rates were 
measured to be $c=(8-11)\cdot10^{-16}$ cm$^{-2}$, compatible for both producers. These results were confirmed in charge collection 
measurements where a similar degradation of the collected charge with fluence was measured for both. The removal rate 
was found to be independent of thickness which points to the deactivation of boron as the main reason 
for the effective acceptor removal.

Thin LGADs retain larger charge collection than standard PIN diodes up to fluences of $\Phi_{eq}\leq 2\cdot 10^{15}$ cm$^{-2}$. The residual concentration
of acceptors though insufficient to provide gain after the depletion of the p$^+$ layer, nevertheless leads at very high bias voltages 
to electric fields required for impact ionization. The devices with a larger 
initial doping concentration of the p$^+$ layer retain gain and moderate voltages up to larger fluences. Beyond $\Phi_{eq}>2\cdot 10^{15}$ cm$^{-2}$ the behavior of LGADs becomes similar 
to that of standard PIN diodes. At highest biases the multiplication takes place in the bulk due to radiation induced deep acceptors, but this mode of operation is limited to voltages close to the breakdown voltage. 

The leakage current follows the prediction given by the bulk generation current and the gain.

\section*{Acknowledgment}

Part of this work has been financed by the Spanish Ministry of Economy and Competitiveness through the Particle Physics National Program (FPA2015-69260-C3-3-R and FPA2014-55295-C3-2-R), by the European Union’s Horizon 2020 Research and Innovation funding program, under Grant Agreement no. 654168 (AIDA-2020).

\end{document}